\documentclass[pra,aps,twocolumn,superscriptaddress]{revtex4-1}
\usepackage[T1]{fontenc}

\usepackage{amsmath,amssymb}
\usepackage{amsfonts}
\usepackage{epsfig,pstricks,graphicx}
\usepackage{bm}
\usepackage{physics}

\definecolor{armygreen}{rgb}{0.29, 0.33, 0.13}
\usepackage{color}
\usepackage{array}

\newcommand{\av}[1]{\left \langle #1 \right\rangle}
\newcommand{\avH}[1]{\left \langle #1 \right\rangle_{HV}}

\usepackage[normalem]{ulem} % either use this (simple) or
\usepackage{soul} % use this (many fancier options)

\begin{document}

\title{Can single photon excitation of two spatially separated modes lead to a violation of Bell inequality via homodyne measurements?}
 
\author{Tamoghna Das}

\affiliation{International Centre for Theory of Quantum Technologies, University of Gdańsk, 80-308 Gdańsk, Poland}

\author{Marcin Karczewski}

\affiliation{International Centre for Theory of Quantum Technologies, University of Gdańsk, 80-308 Gdańsk, Poland}

\author{Antonio Mandarino}

\affiliation{International Centre for Theory of Quantum Technologies, University of Gdańsk, 80-308 Gdańsk, Poland}

\author{Marcin Markiewicz}

\email{marcin.markiewicz@ug.edu.pl}

\affiliation{International Centre for Theory of Quantum Technologies, University of Gdańsk, 80-308 Gdańsk, Poland}

\author{Bianka Woloncewicz}

\affiliation{International Centre for Theory of Quantum Technologies, University of Gdańsk, 80-308 Gdańsk, Poland}

\author{Marek \.Zukowski}

\affiliation{International Centre for Theory of Quantum Technologies, University of Gdańsk, 80-308 Gdańsk, Poland}

\begin{abstract}
We reconsider the all-optical homodyne-measurement based   experimental schemes that aim to reveal Bell nonclassicality of a single photon, often termed `nonlocality'. We focus on the schemes  put forward by Tan, Walls and Collett (TWC, 1991) and Hardy (1994). In the light of our previous work the Tan, Walls and Collett setup can be described by a precise local hidden variable model, hence the claimed nonclassicality of this proposal is apparent, whereas the nonclassicality proof proposed by Hardy is impeccable. In this work we resolve the following problem: which feature of the Hardy's approach 
is crucial for its successful confirmation of nonclassicality. The scheme of Hardy differs from the Tan, Walls and Collett setup in two aspects. (i) It introduces a superposition of a single photon excitation with vacuum as the initial state of one of the input modes of a 50-50   beamsplitter, which  creates the superposition state of two separable (exit) modes under investigation.
 In the TWC case we do not have the vacuum component. (ii) In the final measurements Hardy's proposal utilises a varying strengths of the local oscillator fields, whereas in the TWC case they are constant.  In fact the local oscillators in Hardy's scheme are either on or off  (the local setting is specified by the presence or absence of the local auxiliary field). We show that it is the varying strength of the local oscillators, from setting to setting, which is the crucial feature enabling violation of local realism in the Hardy setup, whereas it is not necessary to use initial superposition of a single photon excitation with vacuum as the initial state of the input mode. Neither one needs to operate in the fully on/off detection scheme.  This implies
 that the confirmed Bell nonclassicality in the Hardy-like setup cannot be attributed to the single-photon state alone, but should rather be considered a consequence of its interference with the photons from auxiliary local fields. 
 Neither can it be attributed to the joint state of the single photon excitation and two separate local oscillator modes, as in the experiment this state is measurement setting dependent.
Despite the failure of the Tan, Walls and Collett scheme in proving Bell nonclassicality, we show that their scheme can serve as an entanglement indicator. We reconsider the general problem of transforming Bell inequalities for optical systems  into indicators of mode entanglement, and find a more efficient entanglement indicator based on a Bell operator than theirs.
\end{abstract}

\maketitle

%\tableofcontents

\section{Introduction}
\label{Intro}
The `nonlocality of a single photon', also known as 
`entanglement with vacuum', has long been a subject of controversy \cite{TWC91, Hardy94, Vaidman95, Hessmo04, Enk05, Dunningham07, Heaney11, Jones11, Brask13, Morin13, Fuwa2015, Lee17}. In its basic form, the problem concerns the nature of the state $\frac{1}{\sqrt2}(|0 1\rangle_{b_1 b_2} +|1 0\rangle_{b_1 b_2})$, obtained by casting a photon on a balanced beamsplitter. Here, the notation $|1 0\rangle_{b_1 b_2}$ indicates the presence of a photon in mode $b_1$ and its absence in $b_2$. Although the resulting state  can be considered as mode entangled \cite{Demkowicz15}, it can also be interpreted as a mere superposition  of {\em mode excitations} -- the photon being either here or there. This point of view is supported by writing down the state as $\hat a^\dagger \ket{00}_{b_1b_2}$, where $\hat a^\dagger = \frac{1}{\sqrt{2}}(\hat b^\dagger_1+ \hat b^\dagger_2)$, and $\hat b^\dagger_j$ are photon creation operators for modes $b_j.$  

Thus, one can question whether it can be used to demonstrate Bell nonclassicality -- both on its own, 
or with some additional resources like local auxiliary optical fields.

Many experiments, some feasible, some gedanken, have been proposed to address this fundamental problem. Let us briefly present their three major types. 

The first one originates from a paper by Tan, Walls and Collett (TWC) \cite{TWC91}, 
in which homodyne-based coincidence intensity measurements with weak coherent light as local auxiliary fields were used to violate the Bell-like inequality of \cite{Reid86}.

However, the  Bell-like inequality in \cite{Reid86} does not rest entirely on Bell's assumptions. Because of that, one can question whether the TWC scheme can be used to violate local realism. For instance,  Santos \cite{Santos92} provided an {\em ad hoc} local realistic model for the correlation functions considered by TWC. Recently, this line of critique has been reinforced in \cite{OurModel} by presenting a model that reproduces  all detection events in the TWC experimental proposal for the range of local oscillator strengths for which the paper \cite{TWC91} reported `nonlocality' of the single-photon state.
In the section III B of this work we show that optical Bell inequalities \cite{Zukowski16}, which must hold for any local realistic description,  are not violated for the TWC setup.
Thus this case is closed.
Still,  TWC correlations are interesting in themselves,  and variants of this scheme have been realized experimentally in \cite{Hessmo04} and \cite{Fuwa2015}. In section V we show how to use the CHSH-like inequality of \cite{Reid86} as an entanglement witness.

Another idea was put forward by Hardy \cite{Hardy94}. He considered states  $q\ket{00}_{b_1b_2}+\frac{r}{\sqrt2}(|0 1\rangle_{b_1 b_2} +|1 0\rangle_{b_1 b_2})$, with $q\neq 0$ which can be produced by sending a superposition of a vacuum state and a single photon one on a 50-50 beamsplitter. Hardy investigated four mutually complementary experimental  situations  and  \emph{proved} that their joint local realistic description would contradict the quantum predictions. His setup relied on tunable amplitudes of the auxiliary coherent fields (local oscillators).  Its modification with mixed-state auxiliary states was proposed in \cite{Dunningham07}, and a generalization to multimode initial states in \cite{Heaney11}. 
 
 Banaszek and W\'odkiewicz suggested a measurement setup implementing the displacement of the input field in the optical phase space for the single photon superposition ($q=0$) \cite{Banaszek99}. The settings of the Bell experiment they considered were defined by  turning the displacements on or off.
 Interestingly, their proof of a violation of the Clauser-Horne (CH) inequality \cite{CH74} in their setup  relied on the no-count events. A combination of this approach with homodyne measurements was investigated in \cite{Lee17}, while an adaptation to multimode input states was given in  \cite{Brask13}. Optical displacement was also used in a scheme for a heralded distribution of a single-particle entanglement  \cite{Caspar20}.  We shall not analyze the approach of \cite{Banaszek99} as, first,  their proof of violation of local realism is impeccable, and second, despite the fact that  their scheme involves only all-optical measurements, the employed technique is essentially different from the ones of TWC and Hardy. 
 
 Finally, van Enk \cite{Enk05} observed that the single-photon superposition  can induce  an entangled state of two  atoms in spatially separated traps. In such a case violation of local realism by the pair of entangled atoms is easy to prove, and the entanglement is (in theory) easily detectable (see e.g., the discussion in \cite{Ashhab07a}). However, this puts us effectively back to the two-qubit entanglement, and the problem is not interesting anymore (except from being experimentally challenging). Moreover, note that the atoms used in such schemes should be treated as auxiliary systems.
 Therefore, such an approach will not be discussed here, as we want to discuss situations which involve only quantum optical fields, passive optics and photon number resolving (macroscopic) detectors.

In this work we shall concentrate on homodyne measurements which involve in general arbitrary beamsplitters, and essentially weak local oscillator fields. This includes 
not only the setups of TWC and Hardy, but also the intermediate cases which have not been discussed yet.
We start by analyzing the mechanism behind the occurrence of the spurious nonclassicality, manifested by the violation of the inequality of \cite{Reid86} in the TWC setup.
Further we shall test the ability of inequalities \cite{Zukowski16} to detect violations of local realism in situations which are a kind of a hybrid of the TWC configuration and the Hardy one. That is, we shall show that  the inequalities detect violation of local realism if one admits varying strengths of the local oscillators for different settings in the Bell experiment. It turns out that the approach allows to find genuine violations of local realism for the initial state of the form of TWC (i.e, for $q=0$).

 \subsubsection{Detailed aims and analysis}
In order to test the potentially nonclassical character of the correlations appearing in the schemes of TWC and Hardy, we use the {\em intensity rates} of optical fields as local observables. They were first introduced as a proposal  of normalized Stokes observables in \cite{StokesFirst1, StokesFirst2} and rediscovered in the context of Bell's theorem in \cite{Zukowski16}, further developed in Refs \cite{Zukowski17, Ryu19}. 
Essentially, by an intensity  rate we mean the ratio of measured intensity in the given local detector to the total intensity measured in all local detectors, in a given run of the experiment.
In contrast to the  CHSH Bell-like inequalities of \cite{Reid86} used by TWC, this approach to analyzing optical correlations does not lead to spurious violations. 

The inequalities for intensities of Ref. \cite{Reid86}
involve  additional constraints on local hidden variable description of  intensities passing beamsplitters. The additional constraints are a version of no-enhancement assumption for CH-like inequalities (discussed in \cite{CH74}), or a fair-sampling one in the case of CHSH-like inequalities.
Therefore the inequalities of Ref. \cite{Reid86} cannot be used to test Bell nonclassicality, however, as we show further, they have an interpretation of entanglement indicators. 
The problematic status of the additional assumption in the inequalities of \cite{Reid86} in the context of TWC setup has been pointed out by Hardy \cite{HardyPhD} and Santos \cite{Santos92}.
Nevertheless, because of the seemingly `innocent' naturality of this assumption, the inequalities of \cite{Reid86} can be found in discussions of Bell's theorem in some textbooks on quantum optics. e.g., \cite{walls2007quantum}.
In this work, we show that  Bell inequalities based on rates of intensities \cite{Zukowski16}, which do not invoke any additional constraints, are  not violated. This result is in perfect agreement with the fact that a recently proposed local hidden variable model \cite{OurModel} reproduces all the probabilities of events considered `nonlocal' by TWC in \cite{TWC91}. It shows that the rate-based inequalities are more reliable than the ones of \cite{Reid86}, as they do not lead to spurious violations of Bell nonclassicality. Interestingly, this is the first found example supporting that claim. It complements the previous investigations in which the rate-based inequalities were always more strongly violated in case of genuine nonclassicality \cite{Zukowski16}.  

In contrast with the above, our analysis of the Hardy-like proposal using  the intensity rates approach indicates the presence of nonclassical correlations.
We have confirmed them in case of the initial state being just a single photon ($q=0$). Thus, the crucial change to the TWC setup introduced by Hardy consists in the ability to vary the intensities of the coherent local oscillators by turning them on or off. 

We have also investigated  a transition between the Hardy-like and  TWC setups. Our numerical calculations show that nonclassical correlations can still be obtained when the  amplitudes of auxiliary coherent fields are non-zero for both local measurement settings. This means that turning the auxiliary oscillators off, as proposed by Hardy, is not necessary. However, their amplitudes must still be different for the alternative local settings of the Bell-type experiment. This strongly suggests that the violations of local realism in such an experiment cannot be associated with a unique state. In other words, the full state $\ket{\alpha_1}_{a_1}\frac{1}{\sqrt2}(|0 1\rangle_{b_1 b_2} +|1 0\rangle_{b_1 b_2})\ket{\alpha_2}_{a_2}$ includes local oscillators, which need to change from setting to setting. Thus the experiment involves more than a single state, which is not consistent with the Bell's theorem.

This in turn suggests that the TWC-Hardy correlations result from an interferometric effect based on bosonic indistinguishability between the single photon in question, and the photons of the auxiliary local coherent beams.
Note that this feature of the investigated setup goes beyond the TWC and Hardy cases. It is in fact characteristic to any nonclassicality test relying solely on single-photon superposition, local oscillators,  and passive optical devices, including the scheme of Banaszek and W\'odkiewicz \cite{Banaszek99}.
The necessity of introducing auxiliary systems in such measurement schemes is related to the fact that it is impossible to perform a projection onto a superposition between vacuum and a single photon using only passive optical devices. This was already pointed out by Peres \cite{Peres95} in a comment to the Hardy's work.

An alternative interpretation is that the local oscillators of the homodynes used to detect the violations of local realism {\it induced} by the state $\frac{1}{\sqrt2}(|0 1\rangle_{b_1 b_2} +|1 0\rangle_{b_1 b_2})$
must be treated as a part of the local measuring devices. 
In fact, in such all optical scenario the local oscillators 
can be incorporated in the operational definition of the measurements which are of the POVM class. 

 \subsubsection{Controversies around the topic }

The topic initiated by the TWC paper stirred up a lot of controversy. For instance, Santos suggested that the intensity correlations in  the Tan-Walls-Collett  scheme can  be explained with  local hidden variables (LHV) \cite{Santos92}, and cannot be used to convincingly demonstrate nonclassicality of a single photon. However, his LHV model was  not reproducing the quantum probabilities of all photon count events. Greenberger, Horne and Zeilinger argued \cite{Greenberger95} that Hardy's setup is not single-particle in nature, as analogous correlations can be observed in three-photon interferometric experiments.  Arriving to a similar conclusion, Vaidman pointed out  \cite{Vaidman95} that Hardy's results are inconsistent with the local-realistic Bohm's description of a single-photon \cite{Bohm52}. Hardy counter-argued to both of these objections \cite{Hardy95Rep}, indicating that his proposal uses on average less than a single photon and that the Bohm's model for a single photon is manifestly nonlocal, therefore no conclusion can be drawn from it in the discussed context.

We review the above discussion of the proposals and objections in order to indicate how unclear the subject issue has proved to be and how many contradictory statements were made.

\section{Experimental Setup}

\begin{figure}
	\centering
\includegraphics[width= 1 \columnwidth]{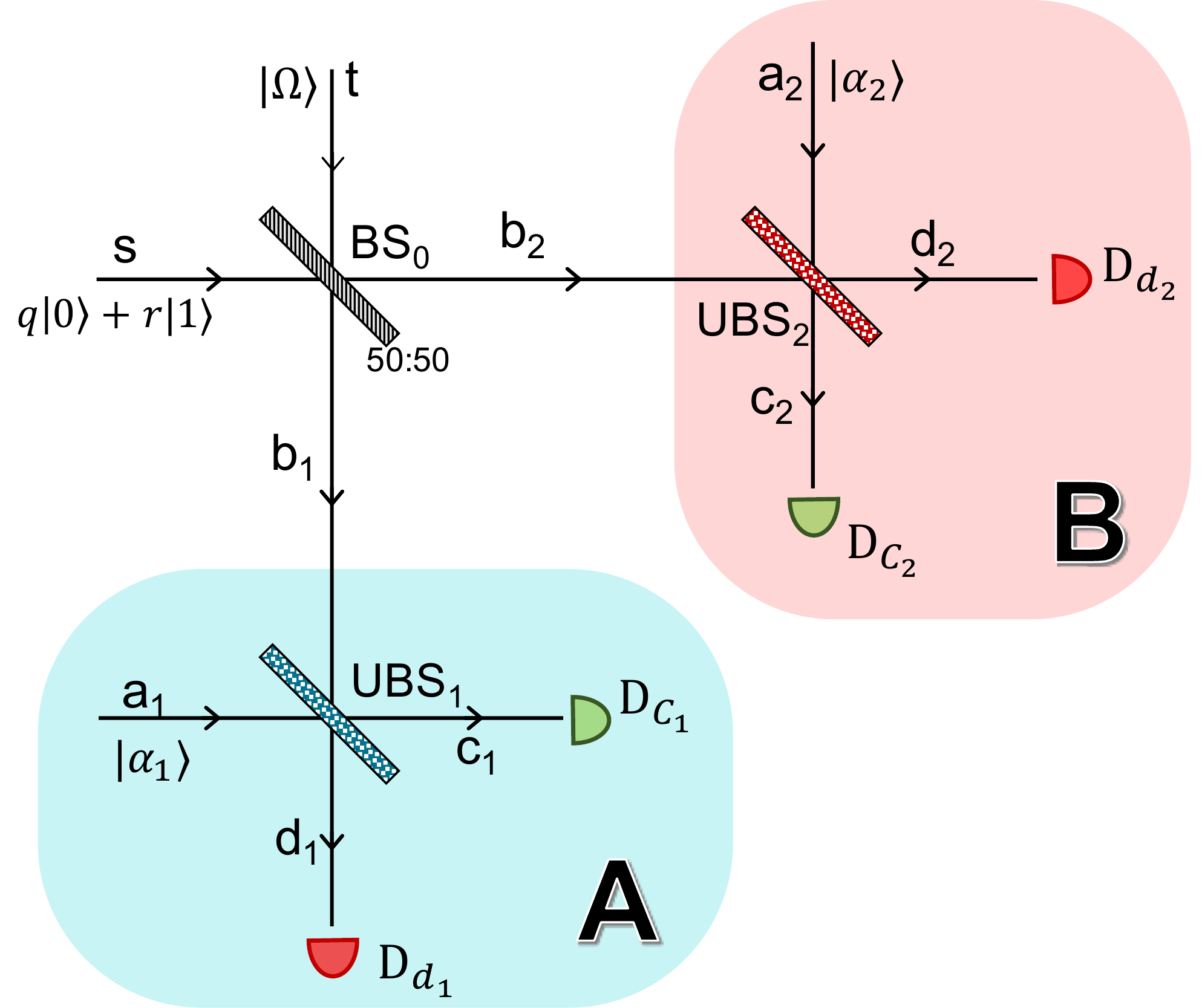}
	\caption{\label{mainSetup}
		Most general schematic representation of  the experimental setup for testing single-photon correlational properties, which we consider here.  In the Tan-Walls-Collett scenario we have $q=0$, and $|\alpha_1|=|\alpha_2|=const$ for all settings in Bell-like experiment. In original Hardy's scenario $q\neq 0$ and $|\alpha_j| = 0$, or $\alpha_j=\frac{i^{(1-j)}r}{q\sqrt{2}}$. Here we consider also intermediate cases, including $\alpha_j$'s of varying absolute values and beamsplitters $U_{BS_j}$ with transmissivity varying from setting to setting.}
\end{figure}

Here, we discuss the basic setup of the TWC and Hardy \textit{gedankenexperiments} and its variations which were studied in the literature, in greater detail.
As depicted in Figure \ref{mainSetup}, 
the experimental configuration consists of three spatially separated beamsplitters $BS_j$ with ${j=0,1,2}$, whose action is in general described by the unitary transformation:
    \begin{equation} \label{SU2Un}
U_{BS}(\chi, \theta) = 
\begin{pmatrix}
\cos \chi & e^{-i \theta}\sin \chi \\
- e^{i \theta}\sin \chi & \cos \chi
\end{pmatrix},   \end{equation}
where $\cos^2  \chi$ is the transmission coefficient of the beamsplitter and $\theta$ is the phase acquired by the reflected beam.
As a beamsplitter is a passive optical device this transformation links the photon  annihilation operators of the incoming beams with annihilation operators of the output beams.
We assume that all photons in the experiment have the same polarization.

An initial state, 
$q \ket{0}_s + r \ket{1}_s$, a superposition of vacuum and single-photon excitation in mode $s$,
impinges on a symmetric beamsplitter BS$_0$  (defined by $\chi = \frac{\pi}{4} $ and $\theta = -\frac{\pi}{2}$). 
 Such will be our notation, $\ket{n}_m$ is a Fock state  of $n$ photons in (spatial) mode $m$.
If the initial state is a single photon, i.e., $q=0$, then it transforms to:
\begin{equation}
	\label{PSI}
	\ket{\psi}_{b_1, b_2} = \frac{1}{\sqrt{2}} \left[ \ket{01}_{b_1, b_2}+ i \ket{10}_{b_1, b_2} \right],
\end{equation}
where $b_1, b_2$, are the output modes of  beamsplitter $BS_0$.

In the TWC scheme the quantum state, given in Eq. \eqref{PSI} is then  shared between two parties Alice and Bob, who perform a homodyne detection.
We assume that the homodyne measurement stations operated by local observers consist of
a $\theta_j$-dependent balanced beamsplitter $BS_j,$ realizing the transformation $U_{BS}(\frac{\pi}{4},\theta_j)$,
%j = 1,2
an auxiliary coherent beam impinging on the remaining input mode of $BS_j$, and two photo-detectors $D_{c_j}$ and $D_{d_j}$, placed in front of the output modes of $BS_j$, for $j = 1,2$, as shown in Figure \ref{mainSetup}. The amplitudes of the coherent beams  $\ket{\alpha}_{a_1}$ and $\ket{\alpha}_{a_2}$ are equal in the case of TWC scheme. For simplicity we assume that $\alpha$ is real. After all that, the total state at the inputs of the two beamsplitters $BS_1$ and $BS_2$, is:
\begin{equation}
|\Psi(\alpha)\rangle=\frac{1}{\sqrt2}\,|\alpha\rangle_{a_1}(|01\rangle_{b_1b_2}+i\,|10\rangle_{b_1b_2})|\alpha\rangle_{a_2}.
\label{BIGPSI}
\end{equation}

The photo-detectors $D_{c_j}$ and $D_{d_j}$ monitor
output modes $\hat c_j$ and $\hat d_j$ of $BS_j$, for $j = 1,2$. These modes  are linked with modes $\hat a_j$ and $\hat b_j$ via symmetric beamsplitter transformation $U_{BS}(\tfrac{\pi}{4},\theta_j)$, that is we have:
\begin{eqnarray}
\label{modeTransf}
\hat c_j&=&\frac{1}{\sqrt2}(\hat a_j+e^{-i\theta_j} \,\hat b_j),\nonumber\\
\hat d_j&=&\frac{1}{\sqrt2}(-e^{i\theta_j}\,\hat a_j+ \hat b_j).
\end{eqnarray}
It is assumed that the detectors measure photon numbers. 
The phases $\theta_1$ and $\theta_2$ in the TWC case are  tunable, and define the local settings in the purported Bell experiment.
Note that in the original TWC setup the tunable phases are attached to the auxiliary coherent beams. While both approaches describe physically equivalent situations, we decided to attach the phases to the beamsplitters, which are part of the local measurement apparatuses, in order to emphasize the fact, that in the TWC setup one deals with \emph{the same initial state} for all measurement settings in the Bell experiment. As we will see, this feature no longer holds in generalisations of the setup.

\subsubsection{Generalizations}

In further discussions we will consider also a generalised scheme with arbitrary tunable  beamsplitters  $BS_1$ and $BS_2$, for which the mode transformation reads:
\begin{equation}
\begin{pmatrix} \label{SU2trans2}
\hat c_j \\
\hat d_j 
\end{pmatrix}    
= U_{BS_j}(\chi_j, \theta_j)
\begin{pmatrix}
\hat a_j \\
\hat b_j 
\end{pmatrix}, 
\end{equation}
where the unitary matrix is given in Eq. (\ref{SU2Un}). Such devices have a realization in the form of Mach-Zehnder interferometers.  One can further generalize the scheme,  and assume that the local auxiliary fields may have different amplitudes for both observers, therefore the overall  state at the entry ports  of the final beamsplitters takes a  more general form:
\begin{equation}
|\Phi(\alpha_1, \alpha_2)\rangle=
\frac{1}{\sqrt2}\,|\alpha_1\rangle_{a_1}(|01\rangle_{b_1b_2}+i\,|10\rangle_{b_1b_2})|\alpha_2\rangle_{a_2}.
\label{BIGPHI}
\end{equation}
In this generalized case the dependence on the \emph{real} parameters $\alpha_1$ and $\alpha_2$ is local with respect to subsystems defined by modes $\{\hat a_1, \hat b_1\}$ and  $\{\hat a_2, \hat b_2\}$, so they can be also used as local settings together with local phases $\theta_1$ and $\theta_2$. However, in contrast to \eqref{BIGPSI}, the initial state is now explicitly setting-dependent.

One can further generalize this to initial states of mode $\hat s$ being superpositions of vacuum and single photon states (i.e $q\neq 0$). In such case the state behind the first beamsplitter is given by:
\begin{equation}
	\label{PSI-HARDY}
	\ket{\psi'}_{b_1, b_2} = q\ket{00}_{b_1, b_2}+\frac{r}{\sqrt{2}} \left[ \ket{01}_{b_1, b_2}+ i \ket{10}_{b_1, b_2} \right].
\end{equation}

\section{Intensities vs. intensity rates in analysis of TWC correlations}
\subsection{Intensity rate measurements allow loophole-free Bell inequalities.}
\label{RateAp}
For the analysis of nonclassicality 
the authors of \cite{TWC91} used Bell-like inequalities derived in \cite{Reid86}. The inequalities involved correlation functions of intensities at pairs of spatially separated detectors, one in Alice's station one in Bob's. They are applicable for the following models:
\begin{equation}
\label{TWCCorr}
 E_T(\theta_1, \theta_2) = \frac{\int d\lambda\rho(\lambda)\prod_{j=1,2}\big (I_{c_j}(\theta_j, \lambda)-I_{d_j}(\theta_j, \lambda)\big)}{\int d\lambda\rho(\lambda)I_1(\lambda)I_2(\lambda)}.
\end{equation}
In the above formula $\lambda$ symbolizes a hidden variable, $\rho(\lambda)$ is  its distribution, $I_{x_j}(\theta_j, \lambda)$ is the hidden variable model of the local intensity  (registered in a detector placed in front of mode $x = c,d$,) at station  $j=1,2$, i.e. respectively A and B of Fig. 1, for the local (phase) setting $\theta_j$. Finally $I_j(\lambda)$ is the total intensity at station $j=1,2$.
However, the inequalities of \cite{Reid86} cannot be used to refute altogether the possiblity of local realistic description of an experiment, as in addition to local realism and `free will', they rely on an additional assumption (which holds in classical optics, but not in e.g. stochastic electrodynamics \cite{Boyer19}). This assumption states  that the total local intensity of light $I_j(\lambda)$, for a given values of a hidden variable $\lambda$ does not depend on the local setting of the measuring device. Namely, the authors of \cite{Reid86}
assumed that:
\begin{equation}
\label{totIntAssumpt}
I_{j}(\lambda)= I_{c_j}(\theta_j, \lambda)+I_{d_j}(\theta_j, \lambda),
\end{equation} 
is {\em independent} of the local value of $\theta_j$. 
A possible violation of inequalities of \cite{Reid86} indicated a failure of either local realism, or `free will', or of the assumption (\ref{totIntAssumpt}). 
However, in an earlier paper \cite{OurModel} we showed an explicit hidden variable model which reproduces {\em exactly} the quantum predictions for the TWC experiment for the {\em whole} range of $\alpha$ for which inequalities of \cite{Reid86} are shown to be violated in \cite{TWC91}. Therefore,  the violation reported in \cite{TWC91} cannot be attributed to Bell's nonclassicality of the state under consideration.

As mentioned above, the inequalities of \cite{TWC91} lead to drastically wrong conclusions about the non-existence of local hidden variable description of TWC experiment. To study this type of questions one must therefore use an alternative approach, which uses Bell's inequalities for field intensities, since they rest only on the assumptions of local realism and `free will'.
The approach given in \cite{Zukowski16}, allows to construct a class of Bell's inequalities having this trait. The Bell's inequalities of \cite{Zukowski16} involve correlations of
functions of the intensities that we will refer to as intensity \emph{rates}.  They are defined as the ratios of the measured intensities in a given local mode to the total intensity measured across all local modes:
\begin{equation}
    \label{RateDef}
    R_{x_j}(\theta_j,\lambda)=\frac{I_{x_j}(\theta_j, \lambda)}{I_{c_j}(\theta_j, \lambda) + I_{d_j}(\theta_j, \lambda)},
\end{equation}
where we assume that $R_{x_j}(\theta_j,\lambda)$ \emph{is assigned the value of $0$ whenever the total intensity in the denominator is equal to $0$}.

Quantum optical observables that describe the rates can be defined as follows \cite{Zukowski16}.
 If one assumes, like in \cite{TWC91}, that the intensity observables of beams are modeled by photon number operators for the given mode, the rate operator is defined by:
\begin{equation}
\hat{\mathrm{R}}_{x_j} = \hat{\Pi}_{c_jd_j}\frac{\hat{n}_{x_j}}{\hat{n}_{c_j}+\hat{n}_{d_j}}
\hat{\Pi}_{c_jd_j},
\label{RATE}
\end{equation}
where  $\hat{n}_{x_j} = \hat{x}^\dagger_j \hat{x}_j$, represents the photon number operator in the $x_j$-th mode of the optical field registered in the detector $D_{x_j}$.  Here,  $\hat{x}^\dagger_j, ~\hat{x}_j$ are
 the  photon creation and annihilation operators of the local output mode  $\hat x_j$, $x=c,d$, of the beamsplitter for observer $j$.
The operators $\hat \Pi_{c_jd_j} = \mathbb{I}_{c_jd_j} - \ketbra{\Omega_{c_jd_j}}, ~j = 1,2$, are projectors onto the subspace spanned by the Fock basis of modes $\hat c_j, \hat d_j$, devoid of the vacuum state.
It is worth mentioning that the two mode vacuum $\ket{\Omega_{c_jd_j}}$ and the operators $\hat{\Pi}_{c_jd_j}$ are invariant under the action of the local mode transformations given in Eqs. (\ref{SU2Un}) and (\ref{SU2trans2}). 
Since the operators $\hat{\mathrm{R}}_{x_j}$ and $\hat{\Pi}_{c_jd_j}\frac{1}{\hat{n}_{c_j}+\hat{n}_{d_j}}
\hat{\Pi}_{c_jd_j}$ commute with each other, depending on needs one can put the rate operator also as follows:
\begin{eqnarray}
\hat{\mathrm{R}}_{x_j}  = \hat{n}_{x_j}\frac{1}{\hat{n}_{tot_j}}
\hat{\Pi}_{c_jd_j}=\hat{\Pi}_{c_jd_j}\frac{1}{\hat{n}_{tot_j}}
\hat{n}_{x_j} \nonumber \\=\hat{\Pi}_{c_jd_j} \frac{1}{\sqrt{\hat{n}_{tot_j}}} \hat{n}_{x_j}\frac{1}{\sqrt{\hat{n}_{tot_j}}}
\hat{\Pi}_{c_jd_j},
\label{RATE2}
\end{eqnarray}
where $\hat{n}_{tot_j}= \hat{n}_{c_j}+\hat{n}_{d_j}$.

\subsection{Homodyne detection with intensity rates and Bell CHSH inequality for rates}
\label{CHSH_non_viol}
The differences of intensity rates \eqref{RateDef}, when averaged over the probability distribution of the LHV, can be used to construct the following correlation functions:
\begin{equation}
\label{RateCorrFun}
E_R(\theta_1, \theta_2) = \avH{\prod_{j=1}^2 \left(R_{c_j}(\theta_j,\lambda)- R_{d_j}(\theta_j,\lambda)\right)},
\end{equation}
depending on the value of the local settings defined by the variable $\theta_j$ for $j=1,2$. 
This correlation function satisfies the CHSH inequality for rates:
\begin{equation}
|E_{R}(\theta_1,\theta_2)+E_{R}(\theta'_1,\theta_2)+E_{R}(\theta_1,\theta'_2)-E_{R}(\theta'_1,\theta'_2)|\leq 2. 
\label{CHSH-eq-OK}
\end{equation}
Note that the above inequality holds regardless of whether the condition \eqref{totIntAssumpt} is satisfied or not, which assures the loophole-free character of the Bell test based on inequality \eqref{CHSH-eq-OK}.
In order to check whether it can be violated in the TWC setup, we introduce quantum implementation of the correlation function \eqref{RateCorrFun} based on the theory of intensity measurements for optical fields with rate operators outlined in Section \ref{RateAp}. It can be applied to write explicitly the correct quantum expectation value of the homodyne  photocurrent difference measured by Alice and Bob: 
\begin{eqnarray}\label{HomoA}
\hat{H}_j(\theta_j) = \hat{R}_{c_j}-\hat{R}_{d_j} 
= \hat{\Pi}_{c_jd_j}\frac{\hat{n}_{c_j} - \hat{n}_{d_j}}{\hat{n}_{c_j}+\hat{n}_{d_j}}\hat{\Pi}_{c_jd_j}, 
\end{eqnarray}
with $j = 1,2$, and the implicit $\theta_j$-dependence is specified via mode transformations \eqref{modeTransf}.
With abuse of notation, we also call  $E_{R}(\theta_1,\,\theta_2)$, the correlation coefficient of the joint homodyne measurement on the initial state $\ket{\Psi(\alpha)}$ \eqref{BIGPSI}, performed by both Alice and Bob:
\begin{eqnarray}
E_{R}(\theta_1,\theta_2)&=&\bra{\Psi(\alpha)} \hat{H}_1(\theta_1) \hat{H}_2(\theta_2) \ket{\Psi(\alpha)}\nonumber\\
&=& A_R\left(\alpha\right)\sin(\theta_1-\theta_2),
\label{Ratecorrelator}
\end{eqnarray} 
where $\theta_1$ and $\theta_2$ are the tunable phases of the local beamsplitters.
The amplitude $A_R(\alpha)$ of the correlation function reads explicitly:
\begin{equation}
    \label{ARamp}
    A_R(\alpha)=\frac{e^{-2\alpha^2}(e^{\alpha^2}-1)^2}{\alpha^2}.
\end{equation}
The correlation function \eqref{Ratecorrelator} depends on the \emph{fixed} initial state parameter  $\alpha$ denoting the amplitude of the auxiliary local field. We can express the correlation function \eqref{Ratecorrelator} as arising from a POVM measurement performed solely on the initial single-photon superposition \eqref{PSI} in modes $b_1$ and $b_2$:
\begin{eqnarray}
E_{R}(\theta_1,\theta_2)&=&\bra{\psi} \hat{\mathcal M}_{b_1}(\alpha, \theta_1) \hat{\mathcal M}_{b_2}(\alpha, \theta_2) \ket{\psi}.\nonumber\\
\label{RatecorrelatorPOVM}
\end{eqnarray} 
The POVM operators $\hat{\mathcal M}_{b_j}(\alpha, \theta_j)$ are explicitly constructed  in Appendix \ref{app:POVMH}. We note that a similar analysis of the homodyne detection in terms of POVM operators has been already performed \cite{Tyc04}, here we extend such an approach to homodyne detection with intensity rate operators \eqref{HomoA}.

\begin{figure}[t]
\includegraphics[width= 1 \columnwidth]{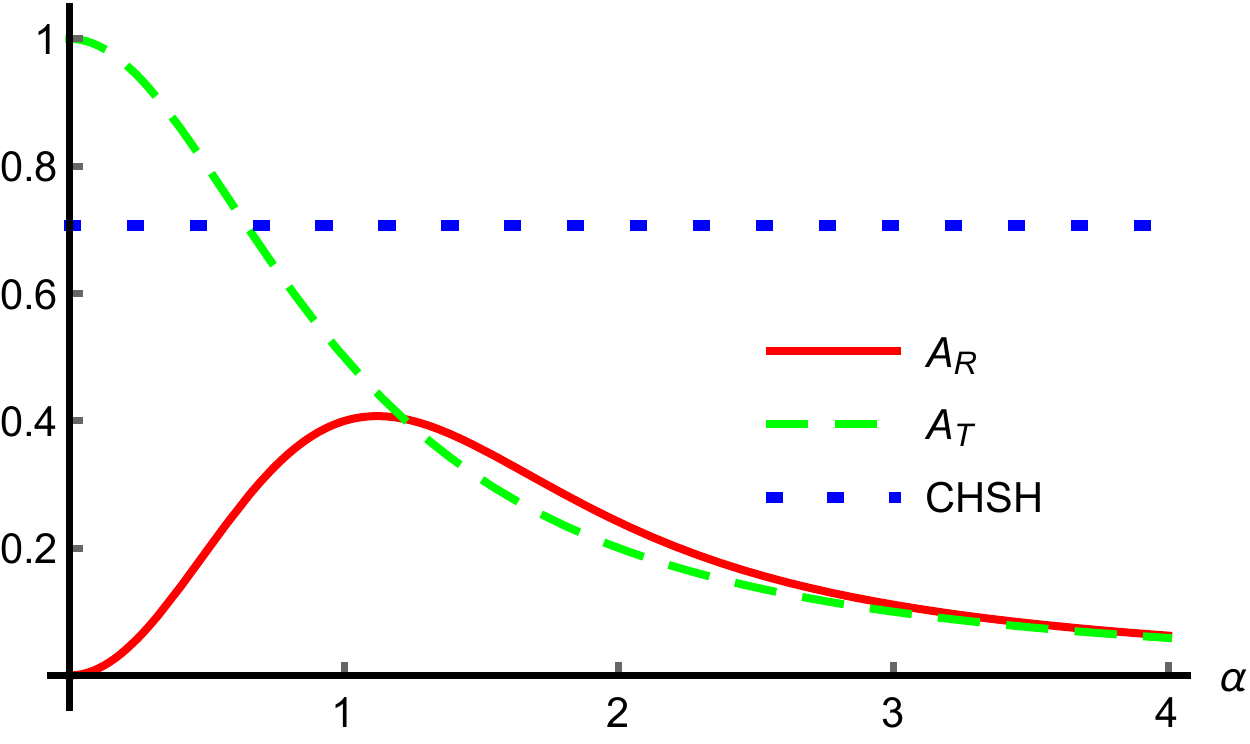}
\caption{\label{fig:amplitude}
Comparison of the amplitudes of correlation functions $E_R(\theta_1,\theta_2)$ and $E_T(\theta_1,\theta_2)$ for different amplitudes $\alpha$ of the auxiliary coherent field. The solid red line  corresponds to the amplitude $A_R(\alpha)$ of $E_R(\theta_1,\theta_2)$, and the dashed green line is for $A_T(\alpha)$. For weak coherent light the amplitude $A_T(\alpha)$ exceeds the threshold needed to violate the CHSH inequality.  }
\end{figure}

Our aim is to  check whether the following CHSH inequality for rates, constructed by substituting \eqref{Ratecorrelator} into \eqref{CHSH-eq-OK}, is violated by the  quantum state given in \eqref{BIGPSI}:
\begin{eqnarray}
\label{CHSHRate}
    &&A_R(\alpha) |\sin(\theta_1-\theta_2) + \sin(\theta'_1-\theta_2) + \sin(\theta_1-\theta_2') \nonumber\\&&- \sin(\theta'_1-\theta'_2)|\leq 2.
\end{eqnarray}
The term depending on the phases of the auxiliary coherent states can reach the maximal value of $2\sqrt{2}$, hence \eqref{CHSHRate} will be violated if  $A_R(\alpha) > \frac{\sqrt 2}{2}$.
The  amplitude $A_{R}(\alpha)$  has been plotted in the Figure \ref{fig:amplitude}, as the solid red line. The blue dotted straight line depicts the value of $\frac{\sqrt{2}}{2}$. The position of the solid red line below the dotted blue straight line, for all values of $\alpha$ shows that the correlation function \eqref{Ratecorrelator} does not lead to any violation of the loophole-free CHSH inequality for rates \eqref{CHSHRate}.

We compare our results with the ones given in \cite{TWC91},  where the following Bell CHSH-like inequality has been considered:
\begin{equation}
|E_{T}(\theta_1,\,\theta_2)+E_{T}(\theta'_1,\,\theta_2)+E_{T}(\theta_1,\,\theta'_2)-E_{T}(\theta'_1,\,\theta'_2)|\leq 2.
\label{WRONGBELL}
\end{equation}  
The correlation functions $E_{T}(\theta_1,\,\theta_2)$ have been already defined in Eq. \eqref{TWCCorr}.
The quantum implementation of the correlators  $E_{T}(\theta_1,\,\theta_2)$ used in \cite{TWC91} reads as follows:
\begin{eqnarray}
E_{T}(\theta_1,\theta_2)
&=&\frac{\langle\Psi(\alpha)|(\hat{n}_{c_1}-\hat{n}_{d_1})(\hat{n}_{c_2}-\hat{n}_{d_2})|\Psi(\alpha)\rangle}{\langle\Psi(\alpha)|(\hat{n}_{c_1}+\hat{n}_{d_1})(\hat{n}_{c_2}+\hat{n}_{d_2})|\Psi(\alpha)\rangle} \nonumber\\
&=& - \frac{\langle\Psi(\alpha)|(\hat a_1 \hat b_1^\dagger - \hat a_1^\dagger \hat b_1)(\hat a_2 \hat b_2^\dagger- \hat a_2^\dagger \hat b_2)|\Psi(\alpha)\rangle}{\langle\Psi(\alpha)|( \hat a_1^\dagger \hat a_1 + \hat b_1^\dagger \hat b_1)( \hat a_2^\dagger \hat a_2 + \hat b_2^\dagger \hat b_2)|\Psi(\alpha)\rangle} \nonumber\\
&=& A_T(\alpha)\sin(\theta_1-\theta_2), \label{TWCcf}
\end{eqnarray}

and the amplitude $A_T(\alpha)$ reads explicitly:
\begin{equation}
    \label{ATamp}
    A_T(\alpha)=\frac{1}{1+\alpha^2}.
\end{equation}
The coefficient $A_T(\alpha)$ as a function of the amplitude $\alpha$ of the auxiliary coherent field  has been plotted in Figure \ref{fig:amplitude}, as the dashed green line. We observe a violation of the inequality  (\ref{WRONGBELL}) for the range $ 0<\alpha^2 < 0.414$. This indication of Bell's nonclassicality is apparent, since in a recent work \cite{OurModel} we have shown an LHV model simulating the discussed correlations. The apparent nonclassicality occurs due to the loophole connected with the additional assumption \eqref{totIntAssumpt}. Therefore the inequality \eqref{WRONGBELL} is not suitable for Bell-type considerations concerning the setup. Nevertheless it can be utilised as an entanglement witness, which is discussed in more details in Section \ref{sec:EW}.

\begin{figure}[t]
\includegraphics[width= 0.85 \columnwidth]{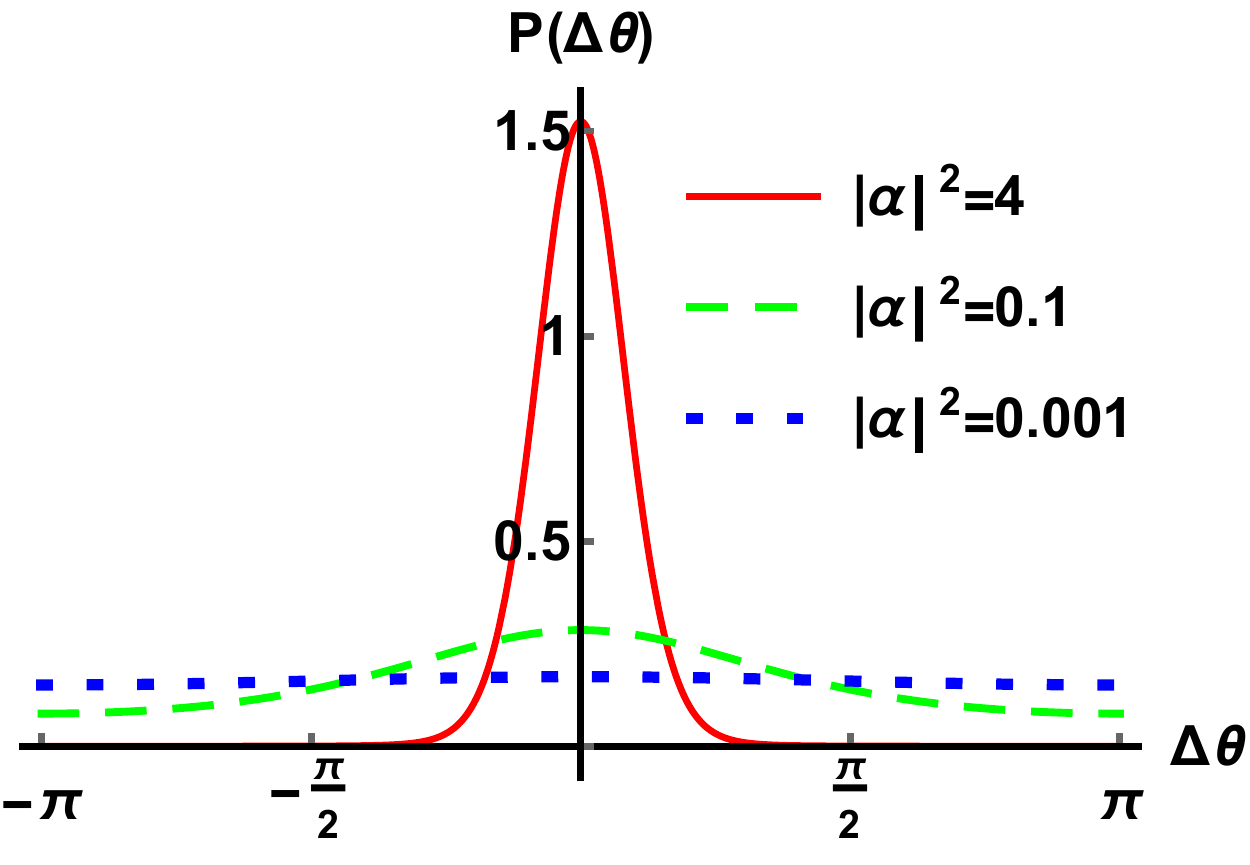}

\caption{\label{fig:pb}
Phase distribution in the Pegg-Barnett formalism. The stronger the coherent light, the narrower the peak of its phase distribution.  }
\end{figure}

\subsection{Classical approximation of the coherent light and the role of phase uncertainty}

In this subsection, we discuss a side but otherwise very interesting issue of a potential trap that may occur when using a classical approximation for the coherent light. Note that all the previous discussion has been done in an approximation-free way.
Below we will see that an interesting paradox occurs in our intensity rate-based approach to the TWC setup if one treats the auxiliary coherent field as a classical beam of light pulses. Such transition can be obtained  under the parametric approximation, that substitutes the annihilation operator of the coherent light with a $c-$number, i.e., its corresponding eigenvalue when acting on the coherent state, namely $a_j\rightarrow \alpha_j$, and similar action when considering the creation operator.
Utilizing the same mode transformations as in \eqref{TWCcf} we obtain the following form of the approximated correlation function \eqref{Ratecorrelator}:
\begin{eqnarray}
 &E^{C}_{R}(\theta_1,\,\theta_2)&\nonumber\\
&= - {}_{b_1,b_2}\langle\psi|\frac{(\alpha_1 \hat b_1^\dagger-\alpha_1^* \hat b_1)(\alpha_2 \hat b_2^\dagger-\alpha_2^* \hat b_2)}{( |\alpha_1|^2+ \hat b_1^\dagger \hat b_1)( |\alpha_2|^2+ \hat b_2^\dagger \hat b_2)}|\psi\rangle_{b_1,b_2}& \nonumber \\
&= \frac{1}{1+ \alpha^2}\sin{(\theta_1-\theta_2)} = A_T(\alpha)\sin(\theta_1-\theta_2)&.
\end{eqnarray} 
This leads to a paradoxical situation in which the intensity rate-based correlation function for a single photon and a classical auxiliary light pulses seems to reveal correlations stronger than before the classical approximation. One can therefore ask, what essential feature of coherent states is lost in this approximation. 

We will now show that the apparent increase of the strength of correlations can be traced back to the phase uncertainty of the coherent states. There are different approaches to defining phase in quantum optics and we use the one introduced by Pegg and Barnett \cite{Pegg}. According to it, in the limit of $\alpha \rightarrow 0$ the phase of a coherent state $|\alpha \rangle$ is distributed as (see also Figure \ref{fig:pb}):
\begin{equation}
P(\Delta \theta)=\frac{1+2\alpha e^{-\alpha^2}\cos{\Delta \theta}}{2\pi}.
\end{equation}
If one takes that into account, the correlation function for weak coherent light taken under the parametric approximation reads as follows when averaged over the phase distribution:
\begin{widetext}
\begin{eqnarray}
 \langle E^{C}_{R}(\theta_1,\theta_2)\rangle_{\Delta\theta_1\Delta\theta_2} &=& \frac{1}{1+\alpha^2}\,\,\iint\limits_{[-\pi,\pi]\times[-\pi,\pi]} P(\Delta\theta_1)P(\Delta\theta_2) \sin{(\theta_1+\Delta\theta_1-\theta_2-\Delta\theta_2)}  d\Delta\theta_1 d\Delta\theta_2\nonumber 
\\
&=& \frac{\alpha^2 e^{-2\alpha^2}}{1+\alpha^2}\sin{(\theta_1-\theta_2)}=A^C_R(\alpha)\sin(\theta_1-\theta_2),
\end{eqnarray} 
\end{widetext}
where we introduced another amplitude of the correlation function:
\begin{equation}
    A_R^C(\alpha)=\frac{\alpha^2 e^{-2\alpha^2}}{1+\alpha^2}.
\end{equation}
After the phase averaging the amplitude of the approximated correlation function $A^C_R(\alpha)$ is strictly lower than $A_R(\alpha)$ and tends to it in the limit of $\alpha\rightarrow 0$. All this discussion shows that one has to be careful when applying the parametric approximation, as ignorance of the phase uncertainty can lead to apparent increase in the strength of correlations.

\section{Violation of Clauser-Horne inequality based on intensity rates}

In the previous section we have shown directly that the Bell-CHSH inequality \eqref{CHSH-eq-OK} based on intensity rates \eqref{RateDef}  is not violated in the TWC setup. This result is concurrent with the existence of an LHV model for the TWC scenario proposed by us in a recent work \cite{OurModel}. The crucial feature of the TWC experiment is the fact that the local oscillators have constant amplitudes. Hence, in order to search for a loophole-free Bell violation within the general single-photon setup (Figure \ref{mainSetup}) we need to relax this assumption, as it was done by Hardy \cite{Hardy94} in his version of the \emph{single-photon nonlocality} test. Contrary to the original Hardy case we do not use the single-photon-with-vacuum superposition as the initial state and keep just the single photon one. On the other hand we introduce one more modification with respect to both TWC and Hardy versions, namely the local beamsplitters $BS_1$ and $BS_2$ in Figure \ref{mainSetup} have now tunable transmissivities, and are specified by formula \eqref{SU2trans2}. In his work \cite{Hardy94} Hardy used Clauser-Horne (CH) inequality \cite{CH74}, as it was better suited to describe his \emph{possibilistic} paradox (meaning that the contradiction occurs with a non-zero probability). 

\subsection{Non-equivalence of CH and CHSH inequalities for optical fields}
The CH inequality in its most general form reads as follows \cite{CH74}:
\begin{eqnarray}
\label{CHgen}
&-1\leq P(A,B)+P(A,B')+P(A',B)-P(A',B')&\nonumber \\
&-P(A)-P(B)\leq 0, &
\end{eqnarray}
where $A,A'$ denote \emph{fixed} outcomes of arbitrary Alice's measurements, and $B,B'$ refer to Bob's ones.
In the case of dichotomic outcomes, namely $A,A',B,B'=\pm 1$, the CH inequality \eqref{CHgen} 
can be written in four versions, each corresponding to a definite choice of $\pm1$ values for all the four outcomes $A,A',B,B'$. Interestingly \emph{each} of these versions is equivalent to the CHSH inequality.
This equivalence rests on the following property of dichotomic measurements: $P(-1|\theta, \lambda)+P(+1|\theta, \lambda)=1.$

The equivalence persists even in the case of imperfect detection if one uses the Garg-Mermin trick \cite{Garg87}. The trick simply leaves them with dichotomic outcomes, despite possible three results, detection in channel +1, in channel -1, or no detection at all. Simply the idea in Ref. \cite{Garg87} is to ascribe value $-1$ also for the non-detection event. 

Note that for qubits the upper bound is often thought to be {\em the} CH inequality, as it can be elegantly derived using geometric concepts \cite{Santos86}. This can be done by noticing that $S(A,B)=P(A)+P(B)-P(A,B)$ is non-negative, $S(A,A)=0$, and it satisfies the triangle inequality. The CH inequality is then the quadrangle inequality.

Any Bell inequality for two, or more qubits can be rewritten into an inequality for intensity rates of fields, via a replacement:
\begin{equation}\label{REWRITE}
P(n_{x_i}=1|\theta_i, \lambda)\rightarrow R_{x_i}(\theta_i|\lambda), 
\end{equation}
where $P(n_{x_i}=1|\theta, \lambda)$ is an LHV probability for the qubit to be registered at $x_i$, while the rates $R_{x_i}$ are defined in Eq. \eqref{RateDef}. For a detailed presentation and justification of this method see \cite{Zukowski16,Ryu19}. 
In the case of the experimental setup discussed throughout this paper the CH inequality for rates can be written in four different versions, depending which pair of output modes we choose from the set $\{(c_1,c_2),(c_1,d_2),(d_1,c_2),(d_1,d_2)\}$. Choosing the pair $(d_1,d_2)$, which coincides with the choice done in Hardy's work \cite{Hardy94}, we obtain the following form:
\begin{eqnarray}\label{CH-RATES}
&-1 \leq \big\langle R_{d_1}R_{d_2}+R_{d_1}R'_{d_2}+R'_{d_1}R_{d_2}-R'_{d_1}R'_{d_2}
&\nonumber \\
&-R_{d_1}-R_{d_2}
\rangle_{HV}\leq 0,&
\end{eqnarray}
in which the unprimed rates correspond to first setting, and the primed ones to the second.

However, it turns out that the CH and CHSH inequalities for rates obtained by application of the procedure \eqref{REWRITE} are no longer equivalent. As a particle can hit only one detector we have 
$\sum_{x_i}P(n_{x_i}=1|\theta, \lambda)=1$. The method (\ref{REWRITE}) works because rates, just like probabilities, have values between zero and one. However, rates of mutually exclusive events, do not have to add up to $1$. We have $\sum_{x_i}R_{x_i}(\theta, \lambda)=1$ or $0.$ 
This {\em prohibits} one to show that the CHSH inequality (\ref{CHSH-eq-OK}) for rates is equivalent to the CH one for rates \eqref{CH-RATES}. Clearly the same conclusion holds for other choices of pairs of output modes for the CH inequality for rates.

\subsubsection{The surprising case of CH inequalities for rates}
In the case of dichotomic outcomes the lower and the upper bound of the CH inequality \eqref{CHgen} are equivalent. Namely by a replacement of $P(A)$ by $1-P(\Tilde{A})$, where $\Tilde{A}$ is an event opposite to $A$, i.e. $P(A)= 1 -P(\tilde{A})$, and subsequent replacements of the form $P(A,B^{(')})=P(B^{(')})-P(\tilde{A},B^{(')})$, one can transform the upper bound of the  inequality \eqref{CHgen} into an inequality of the form of  the lower bound, for events $\{\tilde{A}, A'\}$ for Alice and $\{B,B'\}$ for Bob.

However by performing an analogous replacement of  $R_{d_1}$ by $R_{tot_1}-R_{c_1}$, where $R_{tot_1}=R_{c_1}+R_{d_1}$, which follows the rule (\ref{REWRITE}), utilising the upper bound and transforming it to the lower bound by change of sign of both sides we obtain  the following form of the CH inequality for rates \eqref{CH-RATES}:
\begin{eqnarray}\label{CH-RATES-BAD}
& \big\langle (R_{tot_1}-1)(R_{d_2}+R'_{d_2}) - R_{tot_1}\rangle_{HV}&  \nonumber \\
&\leq \big\langle R_{c_1}R_{d_2}+R_{c_1}R'_{d_2}+R'_{d_1}R'_{d_2}-R'_{d_1}R_{d_2}&\nonumber \\
&-R_{c_1}-R'_{d_2}
\rangle_{HV}.&
\end{eqnarray}
The above inequality is also a CH inequality for rates, where in contrast to the original inequality in \eqref{CH-RATES} the measurement of Alice for the first setting is now performed in mode $c_1$ and the role of primed and unprimed settings for Bob's measurement are swapped.
A glance at the lower bound shows that for $R_{tot_1}=0$ the algebraic expression which is averaged there, reads  $(R_{tot_1}-1)(R_{d_2}+R'_{d_2}) - R_{tot_1}=-R_{d_2}-R'_{d_2}$ and this can reach even the value of $-2$.

Thus the two sides of the CH inequality lead,  via the procedure (\ref{REWRITE}), to two different CH inequalities for rates. 
This explains the phenomenon which we shall see further down. In our analysis we get a violation of only the lower bound of the CH inequality for rates \eqref{CH-RATES}. Surprisingly this inequality is a stronger bound on local realism in the case of rates, than the upper bound. 

\subsection{CH violation in the extended setup}

As already mentioned, in this section we consider  measurement settings that generalize the proposed measurement setup in \cite{TWC91}. Now a single photon ($q = 0$ in our initial state), impinges on a balanced beamsplitter, $BS_0$ and then the output modes interact with auxiliary coherent beams $\ket{\alpha_1}$ and $\ket{\alpha_2}$ through additional two $SU(2)$ beamsplitters $U_{BS_1}(\chi_1,\theta_1)$ and $U_{BS_2}(\chi_2,\theta_2)$, given by the unitary \eqref{SU2Un}. We assume that the amplitudes of the local auxiliary fields $\alpha_1$ and $\alpha_2$ are real.
Following the idea of Hardy setup we treat amplitudes of local fields as a part of measurement settings as well, therefore local settings are specified by vectors:
\begin{eqnarray}
\label{CH_sett00} 
\vec v_1 &=& (\chi_1, \alpha_1, \theta_1),\nonumber\\
\vec v_2 &=& (\chi_2, \alpha_2, \theta_2).
\end{eqnarray}
The  CH inequality for rates specified in the previous section \eqref{CH-RATES} can be expressed in the following convenient form:
\begin{eqnarray}
    \label{CH_ineq}
    -1 \leq && K(\vec v_1, \vec v_2) + K(\vec v_1', \vec v_2) + K(\vec v_1, \vec v'_2) - K(\vec v'_1, \vec v'_2) \nonumber \\ 
    &&- S_1(\vec v_1) - S_2(\vec v_2) \leq 0.
\end{eqnarray}
The correlators $K(\vec v_1, \vec v_2)$ and the local averages $S_j(\vec v_j)$ are defined as follows:
\begin{eqnarray}
    \label{KCorrLHV}
    &&K(\vec v_1, \vec v_2)=\avH{R_{d_1}(\vec v_1)R_{d_2}(\vec v_2)},\nonumber\\
    &&S_j(\vec v_j)=\avH{R_{d_j}(\vec v_j)},
\end{eqnarray}
where the rates $R_{d_j}(\vec v_j)$ are defined as in \eqref{RateDef}, with the only difference that now the hidden local intensities $I_{d_j}(\vec v_j, \lambda)$ explicitly depend on the new local settings $\vec v_j$.

To study the Bell nonclassicality in this extended setup, we replace in (\ref{CH_ineq}) the quantum-mechanical expressions for the two-mode correlators and the local terms:
\begin{eqnarray}
\label{rateCorCH}
 &K(\vec v_1, \vec v_2)&\nonumber\\ &=\bra{\Phi(\alpha_1, \alpha_2)} \hat{R}_{d_1}(\chi_1,\theta_1) \hat{R}_{d_2}(\chi_2,\theta_2) \ket{\Phi(\alpha_1, \alpha_2)},& \nonumber\\
&S_j(\vec v_j) = \bra{\Phi(\alpha_1,\alpha_2)} \hat{R}_{d_j} (\chi_j,\theta_j) \ket{\Phi(\alpha_1, \alpha_2)},&
\end{eqnarray} 
where $\ket{\Phi(\alpha_1, \alpha_2)}$ is a generalised initial state specified in \eqref{BIGPHI}.
Note that we encounter here a peculiar feature, namely that the correlation functions and local averages depend on the settings via parameters of the initial state $\Phi$. If we would like to remove this feature of the setup we have to use POVM operators, and calculate the expressions \eqref{rateCorCH} as follows:
\begin{eqnarray}
&K(\vec v_1, \vec v_2)=\bra{\psi} \hat{\mathcal M}_{b_1}(\vec v_1) \hat{\mathcal M}_{b_2}(\vec v_2) \ket{\psi},\nonumber\\
&S_j(\vec v_j) = \bra{\psi}  \hat{\mathcal M}_{b_j}(\vec v_j) \ket{\psi},
\label{RatecorrelatorPOVMforCH}
\end{eqnarray} 
in which $\ket{\psi}$ denotes the initial single photon superposition \eqref{PSI}, and the POVM operators $\hat{\mathcal M}_{b_j}(\vec v_j)$ are explicitly constructed in Appendix \ref{app:POVMR}.

We choose to detect the transmitted beam $D_{d_j}, ~~j = 1,2$, however we stress that any other choice of the detectors is possible and equivalent conclusions can be addressed.
The terms read as follows: 
\begin{widetext}
\begin{equation}\label{eq:CH1}
\begin{split}
 K(\vec v_1, \vec v_2) = \frac{e^{- \alpha_1^2} e^{- \alpha_2^2}}{2} &\Bigg[ 
\Bigg( \frac{(e^{\alpha_2^2} -1)(1 + e^{\alpha_1^2}(\alpha_1^2-1))}{\alpha_1^2 } + \frac{(e^{\alpha_1^2} -1)(1 + e^{\alpha_2^2}(\alpha_2^2-1))}{\alpha_2^2 } \Bigg) \sin^2 \chi_1 \sin^2 \chi_2  \\
&+ \frac{(e^{\alpha_1^2} -1)(e^{\alpha_2^2} -1)}{\alpha_2^2}  \sin^2 \chi_1 \cos^2 \chi_2
 + \frac{(e^{\alpha_1^2} -1)(e^{\alpha_2^2} -1)}{| \alpha_1|^2} \cos^2 \chi_1 \sin^2 \chi_2 \\
&+ \frac{ (e^{\alpha_1^2} -1)(e^{\alpha_2^2} -1)}{2 \alpha_1 \alpha_2} \sin 2\chi_1 \sin 2\chi_2 \sin( \theta_1- \theta_2) \Bigg], 
 \end{split}
\end{equation}

 \begin{equation}\label{eq:CH2}
  \begin{split}
    S_j(\vec v_j) = 
     \frac{e^{-\alpha_j^2}}{2} \Bigg[ 
    \sin^2 \chi_j \Big((e^{\alpha_j^2} - 1) &+ \frac{1+ e^{\alpha_j^2}(\alpha_j^2-1)}{\alpha_j^2} \Big) + \cos^2 \chi_j \frac{e^{\alpha_j^2} - 1}{\alpha_j^2}  \Bigg].
    \end{split}
 \end{equation}
\end{widetext}
As mentioned we follow the Hardy-like pattern of settings, for which the first setting of each observer corresponds to zero auxiliary field and local beamsplitter acting as identity:
\begin{eqnarray}
\label{CH_sett} 
\vec v_1 &=& (0, 0, 0)\nonumber\\
\vec v_1' &=& (\chi_1', \alpha_1', \theta_1')\nonumber\\
\vec v_2 &=& (0, 0, 0)\nonumber\\
\vec v_2' &=& (\chi_2', \alpha_2', \theta_2').
\end{eqnarray}

We checked by numerical optimization that the double inequality \eqref{CH_ineq} is violated only on the left-hand side and the minimal achievable quantum value reads $-1.0239$. The almost optimal settings for the violation read:
\begin{eqnarray}
\label{CH_sett_fin} 
\vec v_1' &=& \left(\chi_1'=\frac{3\pi}{20}, \alpha_1'=\frac{\sqrt{2}}{2}, \theta_1'=0\right),\nonumber\\
\vec v_2' &=& \left(\chi_2'=\frac{3\pi}{20}, \alpha_2'=\frac{\sqrt{2}}{2}, \theta_2'=-\frac{\pi}{2}\right),
\end{eqnarray}
which corresponds to local beamsplitters with transmissivity about $79\%$ and local coherent fields with average photon number equal to $\tfrac{1}{2}$.

\begin{figure}[t] 
\centering
\includegraphics[width= \columnwidth]{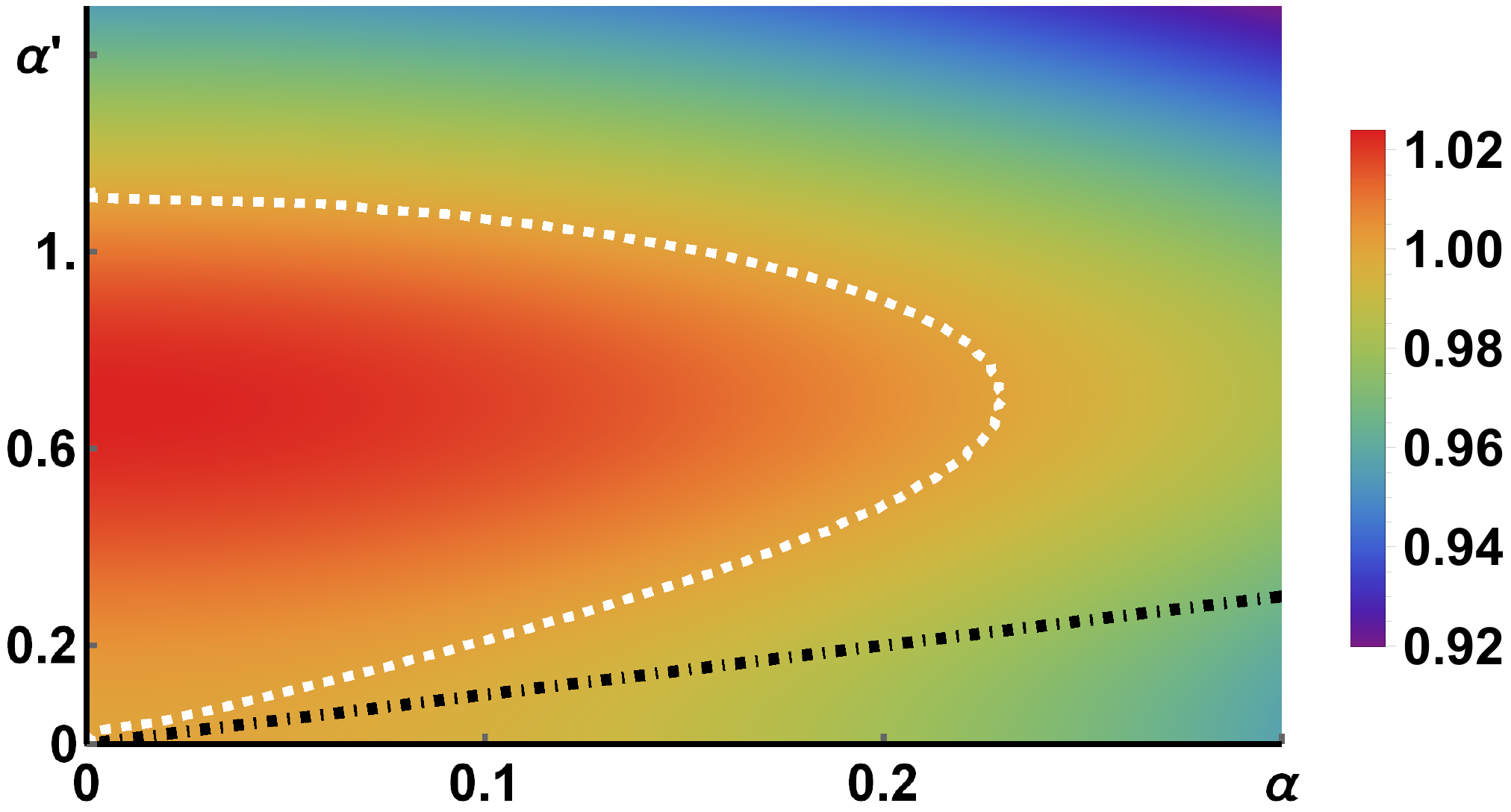}
\caption{Violation of the left-hand-side of the CH inequality \eqref{CH_ineq} as a function of  amplitudes $\alpha, \alpha'$ of the auxiliary fields corresponding to two local settings. The CH-value is sign-flipped for the sake of clarity of the plot. The amplitudes of auxiliary fields \emph{for a given setting} ($\alpha$ or $\alpha'$) are the same for both observers \eqref{CH_sett01}, while they are different for the first  ($\alpha$) and second setting ($\alpha'$).
The black dashed line represents points for which $\alpha=\alpha'$, whereas the white dashed line is an envelope of the violation region with the sign-flipped CH value strictly greater than $1$. Notice that the black-line points lie strictly outside the violation region. This means that it is impossible to obtain the CH violation if we just tune the phases of the local oscillators. The amplitude tuning is necessary for the violation.}
\label{CH_viol_TWC}
\end{figure}

We have also investigated numerically the possibility of violating the CH inequality for rates \eqref{CH_ineq} for the case when for both local settings the auxiliary field has non-zero amplitude, as opposed to \eqref{CH_sett}. We assumed arbitrary parameters for both settings, with the constraint that the amplitudes of auxiliary fields corresponding to a given setting are the same for both observers, which allows for easy graphical representation:
\begin{eqnarray}
\label{CH_sett01} 
\vec v_1 &=& (\chi_1, \alpha, \theta_1),\nonumber\\
\vec v_1' &=& (\chi_1', \alpha', \theta_1'),\nonumber\\
\vec v_2 &=& (\chi_2, \alpha, \theta_2),\nonumber\\
\vec v_2' &=& (\chi_2', \alpha', \theta_2').
\end{eqnarray}
The results of our findings are shown in the Fig. \ref{CH_viol_TWC}, in which we present a plot of a CH-value as a function of the amplitudes $\alpha, \alpha'$ of auxiliary fields corresponding to two settings. The CH-values for each pair of $\alpha$ and $\alpha'$ are optimized over all the remaining parameters in \eqref{CH_sett01}. It turns out that a CH violation still exists for sufficiently small but non-zero values of $\alpha$. On the other hand there is no violation for any value $\alpha=\alpha'$, which is consistent with our previous findings that nonclassical correlations cannot be found for fixed-amplitude auxiliary fields. The situation depicted in the Figure \ref{CH_viol_TWC} shows an intermediate case between the Hardy setup, in which we have on-off tuning of the auxiliary field's amplitudes and the TWC case in which the amplitudes are constant. 

To complete our investigation, we have also examined the original Hardy paradox from the point of view of the intensity rate-based approach. We found that the Hardy setup with intensity rate operators does not lead to a violation of the CH inequality for rates. Our result does not invalidate the Hardy's one, however it indicates a very different nature of the nonclassicality tests based on aggregation of outcomes (like in rate approach) and the ones based on very specific outcomes (like in Hardy case). All the detailed considerations on this topic are presented in a forthcoming publication \cite{3rdPaper}.

\section{Bell operators as entanglement witnesses: intensities vs. rates}
\label{sec:EW}

It is well known that Bell operators can be turned into entanglement witnesses.
Thus, the question arises if the inequality in \cite{Reid86}, used in the TWC paper, could also serve this purpose. As they are not proper Bell inequalities %this
the answer requires a careful analysis. Below we show how to understand the additional condition (\ref{totIntAssumpt}) in this context. Next, we build an optimized entanglement indicator based on the CHSH-like inequality  of \cite{Reid86}. As the method which we employ to build entanglement indicators for fields is also immediately applicable for Bell operators based on intensity rates, we build an analogous entanglement indicator for this case. 

Since the  CHSH-like inequality \eqref{WRONGBELL} seems to be more sensitive to the correlations present in the TWC setup than the CHSH inequality for rates \eqref{CHSH-eq-OK}, it suggests that the   inequality \eqref{WRONGBELL} might be a stronger entanglement indicator than \eqref{CHSH-eq-OK}.
Surprisingly, it turns out the latter inequality leads to an entanglement witness which is more efficient in the case of TWC correlations. That is, it detects entanglement for a broader range of values of $\alpha$.

\subsection{The additional assumption (\ref{totIntAssumpt}) in the case of separable states}
The  original TWC  correlations (with $|\alpha_j|=\alpha= const. $ {for all settings in the experiment}), \cite{TWC91}, do not violate local realism, as shown by the local model presented in \cite{OurModel}. The model covers the whole range of values of $\alpha$ for which TWC obtained violations of the inequality of \cite{Reid86}. As said  earlier,  the inequality of \cite{Reid86} cannot be obtained without the additional assumption (\ref{totIntAssumpt}), which states that for the given value of the hidden variable $\lambda$ the total local intensity predicted by the local hidden variable model does not vary with the measurement settings. Let us assume the model of intensity, which  was used above and leads to the predictions of \cite{TWC91}. It is in the form of number of registered photons (in the given run of the experiment). In such a case the additional assumption transforms into:

\begin{itemize}
    
\item either
\begin{equation}
{n_{{tot}_{i}}(\lambda)} = {n_{c_i}(\theta_i,\lambda)} + {n_{d_i}(\theta_i, \lambda)},
\label{RW_ASSUMP}    
\end{equation}
where ${n_{x_i}(\theta_i,\lambda)}$ is the number of photons predicted by a LHV theory, for the given $\lambda$, to be detected at detector $x_i$, under local setting $\theta_i$. The total number to be detected, in this case, at both outputs ${n_{{tot}_{i}}(\lambda)}$ is assumed to be independent of the local setting. Such must be the form of the condition (\ref{totIntAssumpt}),  if we assume the numbers of photons to be detected are {\em predetermined} by the model for each value of $\lambda$.
\item 
or, for {\em stochastic} local hidden variable models, the condition (\ref{totIntAssumpt}) has to be put as 

\begin{equation}
\langle{{n_{{tot}_{i}}}\rangle (\lambda)} = \langle{n_{c_i}(\theta_i)\rangle(\lambda)} + \langle {n_{d_i}(\theta_i)\rangle ( \lambda)},
\label{RW_ASSUMP-STOCH} 
\end{equation}
where $\langle{n_{x_i}(\theta_i)\rangle(\lambda)}=\sum_{n_{x_i}=1}^\infty n_{x_i}P(n_{x_i}|\theta_i, \lambda)$ for $x = c,d$.
\end{itemize}

Note that condition (\ref{RW_ASSUMP-STOCH}) {\em does not imply} (\ref{RW_ASSUMP}).

We shall show below that in the case of hidden quantum product state model (i.e., for separable states), which is a subclass of LHV models, the condition of \cite{Reid86} should be understood as  (\ref{RW_ASSUMP-STOCH}).

\subsubsection{Condition (\ref{RW_ASSUMP-STOCH}) for separable states}
As already discussed, assumption (\ref{RW_ASSUMP}) imposes an additional constraint on the structure of local hidden variable model.
Still, a local hidden product (quantum) state model is a form of local hidden variable model, which we get when we consider a separable quantum state as the one that causes the correlations. In such a case condition 
(\ref{RW_ASSUMP-STOCH}) definitely holds.

This suggests that the inequality of \cite{Reid86} can be an entanglement witness.

Consider a density matrix describing separable states for optical fields, for the operational scenario which we consider here. They are always a  convex combination of pure ``product'' states (inverted commas are for field theories, see below).
We shall use now $\lambda$ as an index related with any such element of the convex combination, and the weights of the convex combination we shall denote, as earlier, by $\rho_{\lambda}$.
An arbitrary separable state for the optical fields of the considered experiments reads
\begin{eqnarray}
&\rho^{1,2}_{sep}  &\nonumber\\
&=\int d\lambda \rho_{\lambda} f^{\dagger}_\lambda(\hat a^{}_1, \hat b_1)g^\dagger _\lambda(\hat a^{}_2, \hat b_2)\ketbra{\Omega}
f_\lambda(\hat a_1, \hat b_1)g_\lambda(\hat a_2,\hat b_2),&\nonumber\\
\label{RHOSEP}
\end{eqnarray}
where $f^{\dagger}_\lambda(\hat a^{}_1, \hat b_1)$ and $g^\dagger _\lambda(\hat a^{}_2, \hat b_2)$ are polynomial functions of annihilation operators acting on modes $\hat a_i$ and $\hat b_i$  of  $i$-th party. Note that for arbitrary local observables $\hat{O}_i(\hat{a}_i,   \hat{a}_i^\dagger, \hat{b}_i, \hat{b}_i^\dagger)$, and for any of the pure states 
$f^{\dagger}_\lambda(\hat a^{}_1, \hat b_1)g^\dagger _\lambda(\hat a^{}_2, \hat b_2)\ket{\Omega}$
 one has $\langle \hat{O}_1 \hat{O}_2\rangle_\lambda = \langle \hat{O}_1 \rangle_\lambda \langle \hat{O}_2\rangle_\lambda $. Thus, this is a proper description of separability for optical fields.
 
The  probability of getting result $n_{x_1}$ for $1$-st party, with setting $\theta_{1}$ for state indexed by $\lambda$ is  given by:
\begin{eqnarray}
&\text{Pr}(n_{x_1}|\theta_1)_\lambda& \nonumber \\& =
\Tr\big[\delta_{(n_{x_1}, \hat{n}_{x_1}(\theta_1))}f^{\dagger}_{\lambda}(\hat a_1^{},\hat b_1)\ketbra{\Omega}f_{\lambda}(\hat a_1, \hat b_1)\big]& \nonumber \\
&=\langle{\delta_{(n_{x_1}, \hat{n}_{x_1}(\theta_1))}}\rangle_\lambda&
\label{SINGLE_AVG_LAMBDA}
\end{eqnarray}
where $\delta_{(n,k)}$ is the Kroenecker's delta function.

A similar formula holds for 2-nd party for results $n_{x_2}$ under setting $\theta_2$
Following formula (\ref{SINGLE_AVG_LAMBDA}) the  joint probability for two parties goes as follows:
\begin{eqnarray}
&\text{Pr}(n_{x_1},n_{x_2}|\theta_1,\theta_2)_{sep}& 
\nonumber \\
&=   
\Tr \bigg[
\delta_{(n_{x_1}, \hat{n}_{x_1}(\theta_1))}\delta_{(n_{x_2}, \hat{n}_{x_2}(\theta_2))}\rho_{sep}^{1,2}
\bigg]& \nonumber \\
&=\int d\lambda \rho_{\lambda} \text{Pr}(n_{x_1}|\theta_1)_\lambda \text{Pr}(n_2|\theta_{2})_\lambda.&
\label{COMP_AVG_LAMBDA}
\end{eqnarray}

Thus we have a typical structure of a local hidden variable model. Further,  assumption (\ref{RW_ASSUMP-STOCH}) 
is not in contradiction with the structure of separable states. In fact we have an operator identity:
\begin{equation}
{\hat{n}}_{{tot}_{i}}= \hat{n}_{c_{i}}(\theta_i)+\hat{n}_{d_i}(\theta_i),
\label{TOTAL-N}
\end{equation}
which holds for any settings $\theta_i$ and we have
 \begin{equation}
\langle{{\hat n_{{tot}_{i}}}\rangle (\lambda)} = \langle{\hat n_{c_i}(\theta_i)\rangle(\lambda)} + \langle {\hat n_{d_i}(\theta_i)\rangle ( \lambda)},
\label{RW_ASSUMP-STOCH-2} 
\end{equation}
 i.e., $\langle{\hat{n}_{{tot}_i}}\rangle_\lambda$ also does not depend on the settings.  Thus indeed the inequality of \cite{Reid86} is an entanglement indicator.

\subsection*{Entanglement indicators, a comparison}
The inequality of \cite{Reid86} in the TWC version \eqref{WRONGBELL} can be put as the following condition for separability
\begin{eqnarray}
&&\langle \delta\hat  n_1(\theta_1)\delta\hat  n_2(\theta_2) + \delta \hat n_1(\theta'_1)\delta \hat n_2(\theta_2) \nonumber \\ \nonumber
&&+ \delta \hat n_1(\theta_1)\delta \hat n_2(\theta'_2) - \delta\hat n_1(\theta'_1)\delta \hat n_2(\theta'_2)\rangle_{sep} \\
&& \leq 2 \langle\hat n_{{tot}_1} \hat n_{{tot}_2}\rangle_{sep},
\label{TWC_EW}    
\end{eqnarray}
where we introduced: $\delta \hat n_i(\theta_i) =\hat n_{c_{i}}(\theta_i) - \hat n_{d_{i}}(\theta_i)$, and similarly  for $\theta_i'$.
However, the right hand side of this  condition was derived following the additional assumption on local hidden variable models. For separable states this can be lowered, as they form a further constrained set of local hidden variable models.

We have to find its upper bound for separable states. This involves also  search for optimal settings. As in the usual case of two qubits optimal violations of the CHSH inequality take place when Alice and Bob use fully complementary settings, we shall give the separability conditions for such a case.
We are going to use an isomorphic mapping from entanglement witnesses for qubits,  to entanglement indicators for quantum optical fields \cite{Zukowski17, Ryu19}. We derive a separability condition based on Bell inequality for qubits and then map it to optical field operators. As we need only two pairs of settings we reduce the Bloch sphere for qubits to a circle and using the standard Pauli matrices we denote the operator basis for each of them by $\vec{\sigma}_j = ( \sigma_j^x, \sigma_j^z)$, $j=1,2$. The optimal separability  condition based on the CHSH inequality can be obtained if we  choose Alice's and Bob's settings to be fully complementary. With each setting we associate a unit Bloch vector. 
 We can put: $\vec a =(1,0) $, $\vec a' =(0,1) $, $\vec b =\frac{1}{\sqrt{2}}(1, 1) $ and $\vec b' = \frac{1}{\sqrt{2}}(1, -1) $. In this notation the LHS of the CHSH inequality  for any pure product state  of two qubits gives the following:
\begin{eqnarray}
    \label{QUBCOND0}
&&\left\langle \vec a \cdot \vec{\sigma_1}\otimes(\vec b +\vec b' )\cdot\vec{\sigma_2} + \vec a' \cdot\vec{\sigma_1}\otimes(\vec b -\vec b' )\cdot\vec{\sigma_2}\right\rangle_{prod}
\nonumber\\
&&\leq \sqrt{2}
\left(\av{\sigma_1^x}_{prod} \av{\sigma_2^x}_{prod} + \av{\sigma_1^z}_{prod} \av{\sigma_2^z}_{prod}\right)
\leq \sqrt{2}, \nonumber \\
 \end{eqnarray}
because $ \langle\sigma^x_j\rangle^2 +\langle \sigma^z_j\rangle^2 \leq 1$, where  $j=1,2$, for any qubit state. 
Condition (\ref{QUBCOND0}) holds as well for all mixed separable states of two qubits as they are convex combinations of product states. It holds also for arbitrary orthogonal pairs $\vec{a}$, $\vec{a}'$ and $\vec{b}$, $\vec{b}'$. Its final form as an entanglement witness thus reads:
\begin{eqnarray}
    \label{QUBCOND}
&&\langle \sqrt{2}\sigma_1^0\otimes\sigma_2^0-[\vec a \cdot \vec{\sigma}_1\otimes(\vec b +\vec b' )\cdot\vec{\sigma}_2 \nonumber\\
&&+ \vec a' \cdot\vec{\sigma}_1\otimes(\vec b -\vec b' )\cdot\vec{\sigma}_2]\rangle_{sep}\geq 0,
\end{eqnarray}
 where we have introduced identity matrices $\sigma^0$.

A method of finding an  isomorphic entanglement witness given in \cite{Ryu19} is based on replacements $\sigma_\mu \rightarrow \hat{S}_\mu$, where the latter is a quantum Stokes operator of optical field ($\mu=0,1,2,3$). In the case of the TWC interferometric configuration $\hat{S}_0$ can be replaced by the total intensity impinging on the detectors behind the local interferometer, here modelled by  $\hat{n}_{tot_i}$, and
$\hat{S}_1$ with $\hat{S}_2$ by respectively $\delta \hat{n}_i(\theta_i) =\hat{ n}_{c_{i}}(\theta_i) - \hat{n}_{d_{i}}(\theta_i)$ and $\delta \hat{n}_i(\theta_i+\frac{\pi}{2})$. Thus the quantum optical analog for the separability condition for two qubits is given, via the isomorphism of \cite{Ryu19}, in the following form:

\begin{eqnarray} 
\label{TWC_EW_BETTER}
&&\big\langle \sqrt{2}\hat{n}_{tot_1}\hat{n}_{tot_2}-[\delta \hat{n}_1(\theta_1)\delta \hat{n}_2(\theta_2)+\delta \hat{n}_1(\theta_1)\delta \hat{n}_2(\theta_2+\tfrac{\pi}{2})\nonumber\\
&&+\delta \hat{n}_1(\theta_1
+\tfrac{\pi}{2})\delta \hat{n}_2(\theta_2)-\delta \hat{n}_1(\theta_1+\tfrac{\pi}{2})\delta \hat{n}_2(\theta_2+\tfrac{\pi}{2})]\big\rangle_{sep}\nonumber\\&& \geq 0.
\end{eqnarray} 

The above condition (\ref{TWC_EW_BETTER}) can be also transformed into the one with rates instead of intensities. This can be achieved using another  isomorphism presented in \cite{Ryu19}:
\begin{eqnarray}
\label{TWC_EW_BETTER_RATES}
&&\big\langle \sqrt{2}\hat \Pi_{1}\hat \Pi_{2}-
 [ \delta \hat R_1(\theta_1)\delta \hat R_2(\theta_2) +\delta\hat R_1(\theta_1+\tfrac{\pi}{2})\delta \hat R_2(\theta_2)\nonumber\\
 &&+\delta\hat R_1(\theta_1)\delta \hat R_2(\theta_2+\tfrac{\pi}{2}) -\delta \hat R_1(\theta_1+\tfrac{\pi}{2})\delta \hat R_2(\theta_2+\tfrac{\pi}{2})]\big\rangle_{sep}\nonumber\\&& \geq 0,
\end{eqnarray}
where 
$\delta \hat R_i(\theta_i) =\hat R_{c_{i}}(\theta_i) - \hat R_{d_{i}}(\theta_i)$, and the intensity rate operators $\hat R_{x_i}(\theta_i)$ are defined in Eq. \eqref{RATE}.

\begin{figure}[t]
\centering
\includegraphics[width= 1.0 \columnwidth]{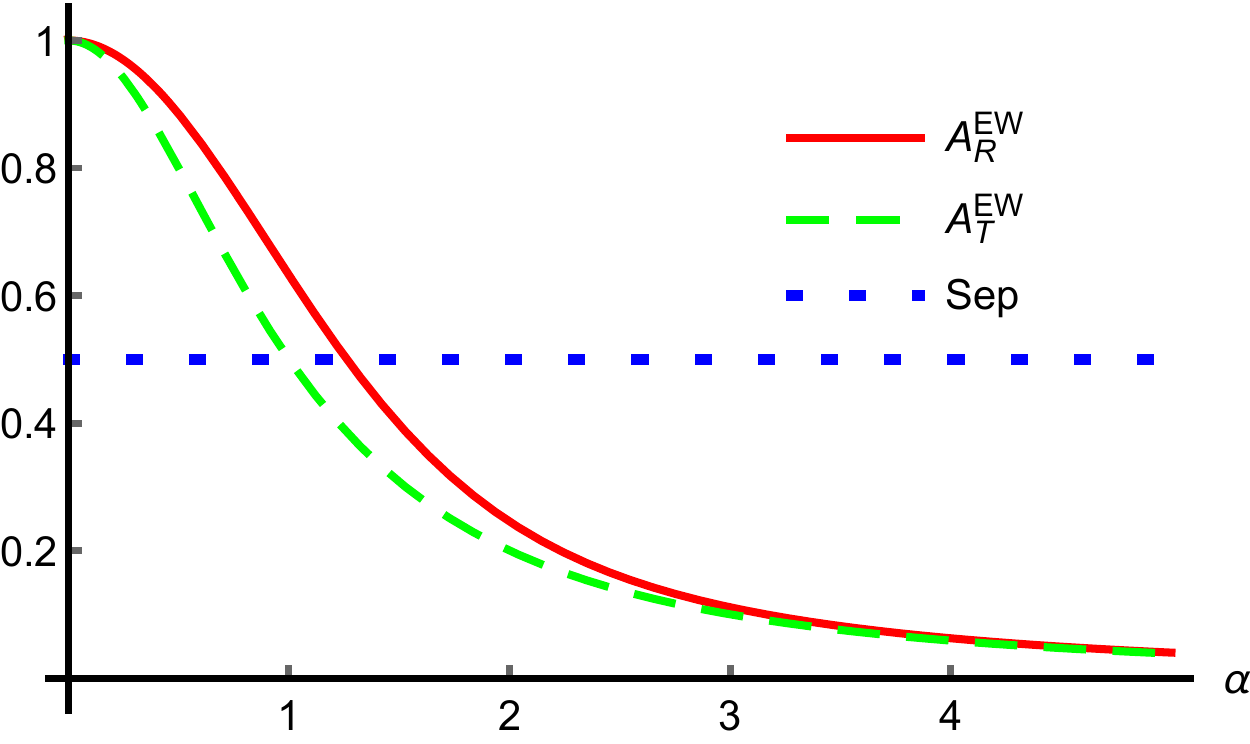}
\caption{ Comparison of amplitudes $A(\alpha)$  in function  of $\alpha$ for CHSH based  entanglement  conditions for rates (\ref{TWC_EW_BETTER}) and  intensities (\ref{TWC_EW_BETTER_RATES}). If $A(\alpha)$ exceeds $\frac{1}{2}$ we observe violation. }
\label{fig:TWCEW}
\end{figure}

Witnesses (\ref{TWC_EW_BETTER})  and (\ref{TWC_EW_BETTER_RATES}) can reveal mode entanglement of the initial state $\ket{\Psi(\alpha)}$ (\ref{BIGPSI}) in a certain range of the coherent field amplitudes $\alpha$. Analogously to (\ref{Ratecorrelator})  we introduce  $A_R^{EW}(\alpha)$  that stands for the amplitude of the correlation function:
\begin{equation}
    \frac{\langle \Psi(\alpha)|\delta \hat R_1(\theta_1)\delta \hat R_2(\theta_2)|\Psi(\alpha)\rangle}{\langle\Psi(\alpha)|\hat \Pi_1 \hat \Pi_2|\Psi(\alpha)\rangle}
    =A_R^{EW}(\alpha)\sin(\theta_1-\theta_2), 
\end{equation}
where the amplitude is given by
\begin{equation}
A_R^{EW}(\alpha)= \frac{e^{- \alpha^2}(e^{ \alpha^2} - 1)}{\alpha^2}. 
\end{equation}
Condition (\ref{TWC_EW_BETTER_RATES})  for $\ket{\Psi}$ can be put as follows:
\begin{eqnarray}
    \label{RATES_PSI}
&A_R^{EW}(\alpha)[ \sin(\theta_1-\theta_2) + 
\sin(\theta'_1-\theta_2) \nonumber \\ &+\sin(\theta_1-\theta'_2)-\sin(\theta'_1-\theta'_2) ] \leq \sqrt{2},
\end{eqnarray}
and its violation  appears if $A_R^{EW}(\alpha) \geq \frac{1}{2}$. The same holds for (\ref{TWC_EW_BETTER})  with the difference that amplitude $A_T^{EW}(\alpha)\equiv A_T(\alpha) = \frac{1}{(1 + \alpha^2)}$ replaces $A_R^{EW}(\alpha)$.  
We present entanglement detection with   (\ref{TWC_EW_BETTER_RATES}) and (\ref{TWC_EW_BETTER}) in terms of values of $A_{R/T}^{EW}(\alpha)$ in  Fig.\ref{fig:TWCEW}.
The CHSH-based entanglement indicator involving rates (\ref{TWC_EW_BETTER_RATES}) is more effective than the one based on CHSH-like inequalities of \cite{Reid86} involving intensities (\ref{TWC_EW_BETTER}). This may seem  surprising, since the CHSH-like inequalities of \cite{Reid86} show (spurious) violations of local realism for the TWC configuration, whereas CHSH-Bell inequality involving rates is  not violated for any $\alpha$. This fact shows the advantage of the rate-based approach over the intensity-based approach : it gives no spurious Bell violations and nevertheless it is more effective as an entanglement indicator.

Of course the presented entanglement indicators have two interpretations, within the context of TWC correlations. Either they detect entanglement of the state:
\begin{equation}
\frac{1}{\sqrt2}\,|\alpha \rangle_{a_1}(|01\rangle_{b_1b_2}+i\,|10\rangle_{b_1b_2})|\alpha \rangle_{a_2}
\label{BIGPSI-22}
\end{equation}
with von Neumann measurements, 
or entanglement of 
\begin{equation}
\frac{1}{\sqrt2}(|01\rangle_{b_1b_2}+i\,|10\rangle_{b_1b_2})
\label{BIGPSI-2222}
\end{equation}
with optical homodyne based POVM measurements. In the second case the settings may depend also on varying moduli of values of local $\alpha$'s.

\section{Discussion and Conclusions}

In our work we reconsidered two prototypical experimental scenarios aiming at demonstration of a single photon nonclassicality, namely the proposal of Tan, Walls and Collett (TWC) \cite{TWC91}, and the one of Hardy \cite{Hardy94}. We have shown that if one applies an intensity-rate-based approach, without using additional assumption (\ref{totIntAssumpt}), the associated Bell-CHSH inequality for rates is not violated in the case of TWC correlations.
This finding is concurrent with our previous work \cite{OurModel}, which shows a failure of TWC setup in demonstrating nonclassicality by providing a precise local hidden variable model for the setup. 
Nevertheless, we have shown that the original TWC analysis can be still used as an entanglement indicator, however a weaker one than the one based on intensity rate approach. These facts are a strong premise towards treating the intensity rate-based approach to analysis of correlations of the optical fields as more appropriate from both: Bell-type perspective (as it is free of a certain loophole) and an entanglement detection perspective (as it gives rise to more robust entanglement indicators). As a side discussion we pointed out possible threats concerning the application of a parametric approximation to a coherent light. Namely if one applies such an approximation at the same time ignoring the phase uncertainty,  spurious increases of the strength of correlations may arise. 

Our intensity rate-based analysis of Hardy-like scenario leads to several conclusions. First, we point out that in this approach the CH inequality for rates is strictly stronger than the CHSH one (for rates). We have shown a violation of the CH inequality for rates in Hardy-like scenario, which includes tunable amplitudes of the coherent local oscillator auxiliary fields, but also (in contrast to the original Hardy analysis) local beamsplitters with tunable transmissivities. Our approach involves modification of two further features of the Hardy scheme, namely we use just a single photon initial state as in the TWC approach ($q=0$ in the Figure \ref{mainSetup}) and do not limit the settings specified by the local oscillator strengths to the on/off detection mode. This indicates that the crucial aspect of the Hardy-like schemes which enables violation of the local realism is the tunability of the strengths of the local auxiliary fields, which need to differ from setting to setting.

 Let us now discuss the physical interpretation of 
 these observations. First, notice that the Hardy-like scenario possesses a peculiar feature untypical for Bell tests, namely that the entire initial state of the interferometric setup is different for different measurement settings. To some extent one can overcome this interpretational difficulty by treating presence of auxiliary fields as an optical implementation of a positive-operator-value measurement (POVM).
 We show analytical forms of such POVMs in Appendix \ref{app:POVMH} and Appendix \ref{app:POVMR}. However even such a treatment does not dispel doubts  whether nonclassical correlations found in Hardy-like scenarios can be attributed solely to the single photon (excitation) state. Despite several objections discussed earlier majority of works devoted to this issue states that the sole source of correlations in such experiments is the single photon superposition. Our reconsideration of the mechanism of the nonclassical correlations in the setup indicates that such a conclusion is unfounded due to a very subtle issue, which goes beyond the duality between mode and particle entanglement in interferometric setups (see e.g. discussion in \cite{Demkowicz15}).
Namely, the source of observed correlations is the quantum interference, which occurs if and only if a given detected multiphoton event could have been realised in two or more genuinely \emph{indistinguishable} ways. The quantum interference observed in Hardy-like schemes is based on indistinguishability of processes corresponding to detection of photons coming from the signal state and the local oscillator fields. At the detection stage due to mentioned interference all  photons must be treated on equal footing, hence the distinction between the input (signal) photons and the local oscillator ones disappears. 

Based on our findings, we argue that the resource behind observed nonclassical correlations in the TWC-Hardy experiments is  quantum interference due to the indistinguishability of photons originating from different sources. This means that the `nonlocality of a single photon' shares the same interpretation as the other profound interferometric nonclassical  phenomena, like photonic entanglement swapping \cite{Zukowski93}, or nonclassical interference  for pairs of photons originating from independent sources \cite{Yurke92, YuSt92, Kaltenbaek06, Blasiak19}.

\section*{Acknowledgements}
We thank \L{}ukasz Rudnicki for comprehensive discussions on the topic of mode entanglement.
The work is part of the ICTQT IRAP (MAB) project of FNP, co-financed by structural funds of EU. MK acknowledges support by the Foundation for Polish Science through the START scholarship.
AM acknowledges support by National Research Center through the grant MINIATURA  DEC-2020/04/X/ST2/01794.

\bibliography{SinglePhoton}
 
\onecolumngrid

\appendix 

\section{Effective POVM operators for CHSH inequality rates}
\label{app:POVMH}
We obtained the correlation coefficient (see Eq. (\ref{Ratecorrelator})) for local homodyne-type measurements performed on the linear superposition of a single photon and vacuum, $\ket{\psi}_{b_1,b_2}=\frac{1}{\sqrt{2}}(\ket{01}+i \ket{10})_{b_1,b_2}$, in terms of intensity rates in mode $\hat c_j$ or $\hat d_j$ 
in a symmetric beamsplitter transformation $U_{BS_j}(\frac{\pi}{4}, \theta_j)$, where an auxiliary coherent field $\ket{\alpha }_{a_j}$, impinges on the remaining input of the beamsplitter for $j =1, 2$. In this section our aim is to figure out the form of the positive operators (POVMs) acting on the initial state $\ket{\psi}_{b_1,b_2}$, which give rise to an equivalent form of the same correlation function \eqref{RatecorrelatorPOVM}. Suppose $\hat{\cal M}_{b_j}(\alpha, \theta_j)$ is the POVM operator in part of the mode $b_j$, for $j = 1,2$, then we obtain the defining condition for the POVM by demanding equivalence of the two forms of the correlation coefficient (compare Eqs \eqref{Ratecorrelator} and \eqref{RatecorrelatorPOVM}):
\begin{equation}
   \bra{\Psi(\alpha)}\hat{H}_1(\theta_1) \hat{H}_2(\theta_2) \ket{\Psi(\alpha)} =  {}_{b_1b_2}\langle\psi|\hat{\cal M}_{b_1}(\alpha, \theta_1) \hat{\cal M}_{b_2}(\alpha, \theta_2) \ket{\psi}_{b_1b_2},
\end{equation}
where the initial state $\ket{\Psi(\alpha)}$ is defined in Eq. \eqref{BIGPSI}.
Hence, the POVM operator is given by:
\begin{eqnarray}
	   && \hat{\cal M}_{b_j}(\alpha, \theta_j) = {}_{a_j}\langle \alpha| ~\hat{H}_j(\theta_j)\ket{\alpha }_{a_j} \\
	   && = {}_{a_j}\langle \alpha  |~ \hat{\Pi}_{c_jd_j} \frac{\hat{n}_{c_j}-\hat{n}_{d_j}}{\hat{n}_{c_j} + \hat{n}_{d_j}} \hat{\Pi}_{c_jd_j} \ket{\alpha }_{a_j} \\
	 && = {}_{a_j}\langle \alpha  |~ \hat{\Pi}_{a_jb_j}  ~\frac{ e^{i\theta_j}  \hat a_j \hat b_j^\dagger + e^{-i\theta_j}  \hat a_j^\dagger \hat b_j}{\hat a_j^\dagger \hat a_j + \hat b^\dagger_j \hat b_j}  ~  \hat{\Pi}_{a_jb_j} \ket{\alpha }_{a_j}  \\
	    && = e^{-\alpha^2}\sum_{n=0}^\infty \frac{\alpha^{2n+1}}{n!} \sum_{m = 1}^\infty  \frac{\sqrt{m}}{n+m} \left( e^{i\theta_j}  \ket{m}\bra{m-1}_{b_j}  +  e^{-i\theta_j}  \ket{m-1}\bra{m}_{b_j} \right).~~~~ \label{eq:POVM}
	\end{eqnarray}

	\begin{figure}[t]
\centering
\includegraphics[width= 0.6 \columnwidth]{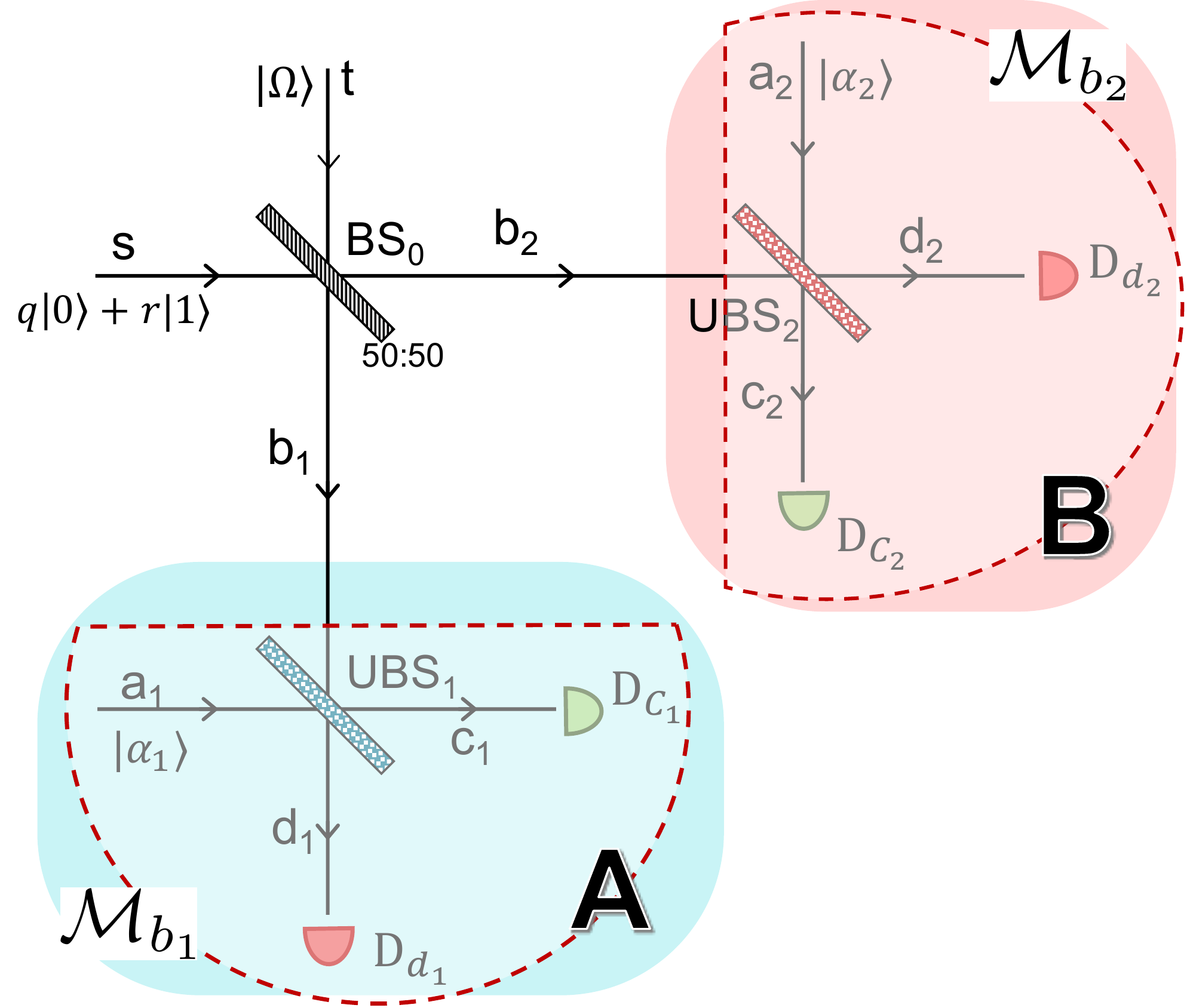}
\caption{Schematic diagram of the effective POVM operators acting on modes $\hat b_1$ and $\hat b_2$ to detect the nonclassicality of a single photon. Average intensity rate $\langle \hat R_{x_j}\rangle$ detected in detector $D_{x_j}$, for $x = c,d, ~j = 1,2$, or the correlation $\langle \hat R_{x_1} \hat R_{x_2}\rangle$ detected in detectors $D_{x_1}$ and $D_{x_2}$ for $x = c,d~~$ on the output modes of the beamsplitters $U_{BSj}, j = 1,2$ along with an auxiliary coherent beam,  can be considered as an effective POVM $\hat{\cal{M}}_{b_j}$ acting on the modes $b_j, ~j = 1,2$. The semicircular shaded region shows that the beamsplitter $U_{BSj}$, the coherent field $\ket{\alpha_j }$, and detection of the the Homodyne  $ \hat{H}_j$ for $j = 1,2$ are equivalent to a POVM $\hat{\cal{M}}_{b_j}(\alpha_j, \theta_j)$, for  $j = 1,2$ is given in Eq. (\ref{eq:POVM}) and the detection of $\hat{R}_{d_j}$, for $j = 1,2$ are equivalent to a POVM $\hat{\cal{M}}_{b_j}(\vec{v_j})$, for  $j = 1,2$ is given in Eq. (\ref{eq:POVM2}).  }
\label{POVM}
\end{figure}

\section{Effective POVM operators for CH inequality and tunable exit beamsplitters}
\label{app:POVMR}
In the previous section, we have calculated the POVM for the balanced beamsplitter and for the homodyne detectors $\hat{H}_j$. Here, we will try to obtain the POVM for the correlation coefficients given in  Eqs. (\ref{rateCorCH}) and (\ref{RatecorrelatorPOVMforCH}) for arbitrary beamsplitters. The construction of the POVM is shown schematically in the Figure \ref{POVM}.

For a beamsplitter with arbitrary transitivity, the output modes are given by (see Eq. (\ref{SU2trans2})):
\begin{eqnarray}
 \hat c_j &=& \cos \chi_j \hat a_j + \sin \chi_j e^{-i\theta_j}  \hat b_j, \\
 \hat d_j &=& -\sin \chi_j e^{i\theta_j}  \hat a_j +  \cos \chi_j \hat b_j.
\end{eqnarray}

Now we want to find out POVM elements ${\hat{\cal M}}_{b_j}(\vec v_j), j = 1,2$, such that the correlation coefficients and the local terms from Eq. \eqref{RatecorrelatorPOVMforCH} read $K(\vec v_1, \vec v_2) = {}_{b_1b_2}\langle \psi| {\hat{\cal M}}_{b_1}(\vec v_1){\hat{\cal M}}_{b_2}(\vec v_2) \ket{\psi}_{b_1b_2}$ and $S_j(\vec v_j) = {}_{b_1b_2}\langle \psi| {\hat{\cal M}}_{b_j} (\vec v_j) \ket{\psi}_{b_1b_2}$, for $j = 1,2$.

By demanding equivalence of the two forms of correlation coefficients in Eqs. \eqref{rateCorCH} and \eqref{RatecorrelatorPOVMforCH} we obtain:
	\begin{eqnarray}
	 && {\hat{\cal M}}_{b_j}(\vec v_j) = {}_{a_j}\langle \alpha_j  |\hat{R}_{d_j}(\chi_j,\theta_j)\ket{\alpha_j }_{a_j} = {}_{a_j}\langle \alpha_j  |~ \hat{\Pi}_{c_jd_j} \frac{\hat{n}_{d_j}}{\hat{n}_{c_j} + \hat{n}_{d_j}} \hat{\Pi}_{c_jd_j} \ket{\alpha_j }_{a_j} \\
	 && = {}_{a_j}\langle \alpha_j  |~ \hat{\Pi}_{a_jb_j}  ~\frac{\hat d_j^\dagger \hat d_j}{\hat a^\dagger_j \hat a_j + \hat b^\dagger_j \hat b_j }  \hat{\Pi}_{a_jb_j} \ket{\alpha_j }_{a_j} \\
	 &&=  {}_{a_j}\langle \alpha_j  |~ \hat{\Pi}_{a_jb_j} \frac{ \sin^2 \chi_j \hat a_j^\dagger \hat a_j + \cos^2 \chi_j \hat b^\dagger_j \hat b_j}{\hat a_j^\dagger \hat a_j + \hat b^\dagger_j \hat b_j} ~  \hat{\Pi}_{a_jb_j} \ket{\alpha_j }_{a_j} \nonumber \\
	 && \hspace{7cm} -\frac 12  \sin 2\chi_j ~{}_{a_j}\langle \alpha_j  |~ \hat{\Pi}_{a_jb_j}  ~\frac{e^{i\theta_j}  \hat a_j \hat b_j^\dagger + e^{-i\theta_j}  \hat a_j^\dagger \hat b_j }{\hat a_j^\dagger \hat a_j + \hat b^\dagger_j \hat b_j} ~ \hat{\Pi}_{a_jb_j} \ket{\alpha_j }_{a_j} \\
	    &&= e^{-\alpha_j^2} \bigg( \cos^2 \chi_j \left(\mathbb{I}_{b_j} - \ket{0}\bra{0}_{b_j}\right) + \sum_{n=1}^\infty \frac{\alpha_j^{2n}}{n!} ~ \frac{\sin^2 \chi_j ~ n + \cos^2 \chi_j ~b_j^\dagger b_j }{n + b_j^\dagger b_j} \nonumber \\ 
	   && \hspace{4cm} -\frac 12  \sin 2\chi_j \sum_{n=0}^\infty \frac{\alpha_j^{2n+1}}{n!} \sum_{m = 1}^\infty \frac{\sqrt{m}}{n+m} \left( e^{i\theta_j}  \ket{m}\bra{m-1}_{b_j}  +  e^{-i\theta_j}  \ket{m-1}\bra{m}_{b_j} \right) \bigg).~~~~ \label{eq:POVM2}
	\end{eqnarray}
	
If beamsplitters $U_{BS1}$ and $U_{BS2}$ have $100\%$ transmissivity, then the above POVM reduces to $\hat{\Pi}_{b_j} = \mathbb{I}_{b_j} - \ket{0}\bra{0}_{b_j}, ~j = 1,2$.

\end{document}